\documentclass[journal]{IEEEtran}
 \pdfoutput=1

\usepackage{graphicx}
\usepackage{epstopdf}
\usepackage{mathrsfs}
\usepackage{amsfonts}
\usepackage{multirow}

\usepackage{cite}
\usepackage[numbers,sort&compress]{natbib}

\ifCLASSINFOpdf
\else
\fi
\usepackage[cmex10]{amsmath}
\usepackage[tight,footnotesize]{subfigure}

\usepackage{cases}
\begin{document}
%
\title{Combined Field Integral Equation Based Theory of Characteristic Mode}
%
%
%

\author{Qi~I.~Dai,
	 Qin S.~Liu,
	 Hui~Gan,
	 Weng~Cho~Chew
}

\maketitle

\begin{abstract}
Conventional electric field integral equation based theory is susceptible to the spurious internal resonance problem when the characteristic modes of closed perfectly conducting objects are computed iteratively. In this paper, we present a combined field integral equation based theory to remove the difficulty of internal resonances in characteristic mode analysis. The electric and magnetic field integral operators are shown to share a common set of non-trivial characteristic pairs (values and modes), leading to a generalized eigenvalue problem which is immune to the internal resonance corruption. Numerical results are presented to validate the proposed formulation. This work may offer efficient solutions to characteristic mode analysis which involves electrically large closed surfaces.

\end{abstract}


%
\IEEEpeerreviewmaketitle

\section{Introduction}
%
%
%
%
\IEEEPARstart{C}{haracteristic} modes (CM) are originally defined by Garbacz as a basis to diagonalize the scattering and perturbation matrices of a conducting object \cite{Garbacz}. The theory was later refined by Harrington and Mautz \cite{Harrington1971,Harrington1971a}, where an electric field integral equation (EFIE) based derivation was formulated. The EFIE based theory for characteristic mode analysis becomes most widely accepted due to its mathematical elegance, reliability and convenience for practical applications. In \cite{Nalbantoglu82}, Nalbanto\u{g}lu adopted the magnetic field integral equation (MFIE) to determine the characteristic modes of simple bodies of revolution.  Unfortunately, this work has not drawn much attention as it appears to provide no advantage over the EFIE based approach.

Theory of Characteristic mode (TCM) was recently popularized in the antenna community by Cabedo-Fabres \cite{Fabres2007} as it has useful applications in antenna shape synthesis, input impedance matching, and radar cross-section control, etc. Although TCM has been shown promising for systematic antenna design, the theoretical  and numerical aspects are relatively less addressed. Characteristic mode analysis requires one to solve a generalized eigenvalue problem as $\bar{\mathbf{X}} \mathbf{J}_n = \lambda_n \bar{\mathbf{R}}   \mathbf{J}_n $, where the dense matrix pair $(\bar{\mathbf{X}},\bar{\mathbf{R}})$ is generated from the EFIE impedance matrix $\bar{\mathbf{Z}}$. When the problem scale is small, one can use Schur decomposition to find all eigenpairs (eigenvalues and eigenvectors) of the generalized eigenvalue problem with a computational complexity of $O(N^3)$, where $N$ is the number of unknowns. This becomes increasingly unaffordable as the problem scale increases. However, in many applications, only $\mathbf{J}_n$ with small $|\lambda_n|$ are desired. Iterative eigensolvers such as the Lanczos and Arnoldi methods can be used for this purpose. For instance, the commercial software FEKO incorporates the ARnoldi PACKage (ARPACK) which implements the Implicitly Restarted Arnoldi Method (IRAM) to compute a few characteristic modes of interest of more complicated geometries \cite{Sorensen,Lehoucq98,Ludick}. Such iterative methods are normally more efficient than the direct eigen-decomposition, since they only require a number of matrix-vector multiplication (MVM) operations, each of which has a computational complexity of $O(N^2)$. The complexity of each MVM can be further reduced to $O(N \log N)$ if a modified multilevel fast multipole algorithm (MLFMA) is employed \cite{Song97,Chew2000,Dai14}.

Usually, one can easily obtain good approximations to the eigenvalues of largest magnitude with Krylov methods such as IRAM \cite{Lehoucq98}. Fast convergence to the desirable spectrum requires one to transform the aforementioned generalized eigenvalue problem to a standard one as $\bar{\mathbf{X}}^{-1}  \bar{\mathbf{R}}  \mathbf{J}_n= \lambda_n^{-1} \mathbf{J}_n$. Hence, a number of $\bar{\mathbf{X}}^{-1}\mathbf{u}$ has to be computed, where $\mathbf{u}$ are arbitrary vectors. In many cases, $\bar{\mathbf{X}}^{-1}\mathbf{u}$ can only be computed iteratively since the direct decomposition of $\bar{\mathbf{X}}$ is not feasible. An equivalent eigenvalue problem for computing characteristic modes is given by $\bar{\mathbf{Z}}^{-1} \bar{\mathbf{R}}  \mathbf{J}_n = (1+i\lambda_n)^{-1} \mathbf{J}_n$, which calls for iterative solutions of EFIEs $\bar{\mathbf{Z}}^{-1}\mathbf{u}$. Apparently, it is favored to solve $\bar{\mathbf{Z}}^{-1}\mathbf{u}$ as there exists efficient techniques to precondition EFIEs. However, if the EFIE based TCM is applied to closed conducting surfaces, $\bar{\mathbf{Z}}$ becomes ill-conditioned when the operating frequency is selected close to frequencies of spurious internal resonances. The internal resonance problem is even more severe for electrically-large objects. 

The difficulty of internal resonances has been well addressed in many radiation and scattering problems, where the combined field integral equation (CFIE) is an effective remedy \cite{ChewBook2009}. Although a full-rank $\bar{\mathbf{Z}}$ can be obtained by combing the electric and magnetic field integral operators, it is not transparent to update the vector $\mathbf{u}$ consistently in the eigenanalysis where characteristic modes are iteratively computed. In this paper, an MFIE based TCM is formulated, which is inspired by the very brief result documented in \cite{Nalbantoglu82}. Then, a CFIE based TCM can be formulated as the EFIE and MFIE based TCMs share the same set of non-trivial characteristic pairs (trivial and spurious ones form the null space of the impedance matrix). Moreover, the corresponding characteristic mode expansion of any excited surface current is obtained. Since EFIE can be preconditioned with a Calder\'{o}n multiplicative preconditioner (CMP) \cite{Andriulli,Contopanagos02,Bagci09}, we can further formulate a CMP-CFIE based TCM, which is free from internal resonances, and can be easily solved using iterative eigensolvers after being transformed to a standard eigenvalue problem. 

%

\section{EFIE Based TCM}
The EFIE for an arbitrarily shaped perfect electric conductor (PEC) object relates the tangential incident field $\mathbf{E}_{inc}$ and the scattered field  $\mathbf{E}_{sca}$ on the PEC surface $S$ as
\begin{equation}
\hat n \times \mathbf{E}_{sca}(\mathbf{r}) = \mathcal{Z}_E(\mathbf{r},\mathbf{r}') \cdot \mathbf {J}(\mathbf{r}') = -\hat n \times \mathbf{E}_{inc}(\mathbf{r}), \quad \mathbf{r}\in S
\end{equation}
where
\begin{equation}
\mathcal{Z}_E(\mathbf{r},\mathbf{r}') \cdot \mathbf {J}(\mathbf{r}') = \hat n \times \mathcal{L}_E (\mathbf{r},\mathbf{r}')\cdot \mathbf {J}(\mathbf{r}')
\end{equation}
In the above, we denote
\begin{equation}
\mathcal{L}_E (\mathbf{r},\mathbf{r}')\cdot \mathbf {J}(\mathbf{r}')  = i k\eta\int_S d\mathbf{r}' \, \overline{\mathbf{G}}(\mathbf{r},\mathbf{r}')\cdot \mathbf {J}(\mathbf{r}'),
\end{equation}
and $\mathbf{J}(\mathbf{r}')$ is the induced surface electric current, $\hat n$ is the unit normal of $S$.  The Green's dyadic is
\begin{equation}
\overline{\mathbf{G}}(\mathbf{r},\mathbf{r}') = \left \lbrack \overline{\mathbf{I}} + \frac{\nabla \nabla}{k^2} \right \rbrack 
g(\mathbf{r},\mathbf{r}')
\end{equation}
with the unit dyad denoted as $\overline{\mathbf{I}}$. The scalar Green's function is 
\begin{equation}\label{SGF}
g(\mathbf{r},\mathbf{r}') = \frac{e^{i k  R}}{4 \pi R } , \qquad R = \vert \mathbf{r} - \mathbf{r}' \vert
\end{equation}
where $\mathbf{r}$ and $\mathbf{r}'$ denote the field and source points, respectively. Moreover, the wavenumber is $k=\omega \sqrt{\mu \varepsilon}$, while the wave impedance is $\eta=\sqrt{\mu/\varepsilon}$, where the permittivity and permeability of the homogeneous medium are denoted by $\varepsilon$ and $\mu$, respectively. 

The most widely adopted TCM is formulated on top of EFIE. As suggested by Harrington and Mautz \cite{Harrington1971a}, the characteristic modes of arbitrary PEC objects can be found by solving the operator eigenvalue problem given by
\begin{equation}\label{E_GEP1}
\mathcal{X}_E(\mathbf{r},\mathbf{r}') \cdot \mathbf {J}_n(\mathbf{r}') = \lambda_n \mathcal{R}_E(\mathbf{r},\mathbf{r}') \cdot \mathbf {J}_n(\mathbf{r}'), \quad \mathbf{r}\in S
\end{equation}
where 
\begin{equation}
\mathcal{R}_E = \frac{1}{2}\left(\mathcal{Z}_E + \mathcal{Z}_E^*\right), \quad \mathcal{X}_E = \frac{1}{2i}\left(\mathcal{Z}_E - \mathcal{Z}_E^*\right)
\end{equation}
with the complex conjugate denoted by $*$, and $\mathbf{J}_n(\mathbf{r}')$ denotes the characteristic currents, and  $\lambda_n$ denotes the characteristic values. An equivalent TCM to (\ref{E_GEP1}) is easily obtained as
\begin{equation}\label{E_GEP2}
\mathcal{Z}_E(\mathbf{r},\mathbf{r}') \cdot \mathbf {J}_n(\mathbf{r}') = (1+i\lambda_n) \mathcal{R}_E(\mathbf{r},\mathbf{r}') \cdot \mathbf {J}_n(\mathbf{r}'), \; \mathbf{r}\in S
\end{equation}

Approximating the characteristic currents with expansion functions $\mathbf{f}_j(\mathbf{r}')$, and weighting the two operator eigenvalue problems  (\ref{E_GEP1}) and (\ref{E_GEP2}) with testing functions $\mathbf{t}_i(\mathbf{r})$, one can obtain two matrix eigenvalue equations as 
\begin{equation}\label{E_DGEP1}
\bar{\mathbf{X}}_E  \mathbf{J}_n = \lambda_n \bar{\mathbf{R}}_E  \mathbf{J}_n 
\end{equation}
and
\begin{equation}\label{E_DGEP2}
\bar{\mathbf{Z}}_E  \mathbf{J}_n = (1+i\lambda_n) \bar{\mathbf{R}}_E  \mathbf{J}_n 
\end{equation}
In the above, $\bar{\mathbf{R}}_E$ and $\bar{\mathbf{X}}_E$ are taken as the real and imaginary parts of the EFIE impedance matrix $\bar{\mathbf{Z}}_E$, respectively, while  $\bar{\mathbf{Z}}_E$ is the matrix representation of the operator $\mathcal{Z}_E$. The eigenvectors $\mathbf{J}_n$ contain basis expansion coefficients as entries. 
For EFIEs, Rao-Wilton-Glisson (RWG) and $\hat n \times$RWG basis functions are chosen as $\mathbf{f}_j$ and $\mathbf{t}_i$, respectively.

Characteristic values $\lambda_n$ is important as $\vert \lambda_n \vert$ indicates the modal behavior. When $\lambda_n=0$, the corresponding $\mathbf{J}_n $ is an externally resonant mode which is efficient in radiating energy. When $\lambda_n>0\, (\lambda_n<0)$, $\mathbf{J}_n$  is an inductive (capacitive) mode which stores predominantly magnetic (electric) energy. When $\vert \lambda_n \vert = \infty$, $\mathbf{J}_n$ corresponds to the internal resonance (trivial and spurious) modes of the closed PEC surface ($\mathcal{Z}_E \cdot \mathbf{J}_n=0$), which has no contribution to radiated or scattered fields.  In this case, $\bar{\mathbf{Z}}_E$, $\bar{\mathbf{R}}_E$ and $\bar{\mathbf{X}}_E$ are rank-deficient, and it is not feasible to convert generalized eigenvalue problems (\ref{E_DGEP1}) and (\ref{E_DGEP2}) to standard ones as
\begin{equation}\label{E_DSEP1}
 \bar{\mathbf{X}}_E^{-1} \bar{\mathbf{R}}_E  \mathbf{J}_n =(\lambda_n)^{-1} \mathbf{J}_n 
\end{equation}
and
\begin{equation}\label{E_DSEP2}
\bar{\mathbf{Z}}_E^{-1}  \bar{\mathbf{R}}_E \mathbf{J}_n = (1+i\lambda_n)^{-1}   \mathbf{J}_n 
\end{equation}
whose desired spectra can be easily found iteratively.

TCM may be formulated using Yaghjian's augmented EFIE (AEFIE) \cite{Yaghjian81}. By demanding that the normal component of the electric flux density $\mathbf{D}(\mathbf{r}) = \varepsilon \mathbf{E}(\mathbf{r})$ be equal to the surface charge density on $S$, or
\begin{equation}
\hat n \cdot \mathbf{E}(\mathbf{r}) - \frac{1}{i\omega\varepsilon}\nabla \cdot \mathbf{J}(\mathbf{r}) = 0, \qquad \mathbf{r} \in S
\end{equation} 
one has
\begin{equation}\label{Za}
\mathcal{L}_E (\mathbf{r},\mathbf{r}') \cdot \mathbf{J}(\mathbf{r}') + \frac{i\eta}{k} \hat n \nabla \cdot \mathbf{J} (\mathbf{r}) = - \mathbf{E}_{inc}(\mathbf{r}), \quad \mathbf{r} \in S
\end{equation}
Taking $\hat n \times$ and $\hat n \cdot$ on both sides of (\ref{Za}) yield two equations with consistent solutions. The resultant matrix equations form an over-determined system whose solution can be solved using the method of least squares. This scheme is immune to the internal resonance corruption.  Such AEFIE based TCM may be constructed as
\begin{equation}\label{aefie_gep}
\begin{aligned}
&\Im m [\mathcal{L}_E(\mathbf{r},\mathbf{r}')] \cdot \mathbf{J}_n(\mathbf{r}') + \frac{i \eta}{k} \hat n \nabla \cdot \mathbf{J}_n (\mathbf{r}) \\
& \hspace{1.5in}= \lambda_n \Re e[\mathcal{L}_E(\mathbf{r},\mathbf{r}')] \cdot \mathbf{J}_n(\mathbf{r}')
\end{aligned}
\end{equation}
with $\mathbf{r} \in S$, which can be solved using a similar discretization method. Also of note is that this scheme may fail when a sphere is studied.
    
\section{MFIE Based TCM}
To obtain an MFIE based TCM, we first consider the characteristic field $\mathbf{E}_n$ radiated by a characteristic current $\mathbf{J}_n$, that is
\begin{equation}\label{J2E}
\nabla \times \nabla \times \mathbf{E}_n - k^2 \mathbf{E}_n = i  k \eta \mathbf{J}_n
\end{equation}    
According to Harrington and Mautz \cite{Harrington1971}, the scattering operator which operates on incoming waves to yield outgoing waves is diagonalized when characteristic fields $\mathbf{E}_n$ are chosen as the basis of outgoing waves, and their complex conjugates $\mathbf{E}_n^*$ as the basis of incoming waves. Based on that $\mathbf{J}_n$ is real (equiphase), one can easily argue that $\mathbf{E}_n+\mathbf{E}_n^*$ is a source-free field using (\ref{J2E}). Hence, if source-free fields are written in a linear superposition form, i.e., $\sum_n a_n(\mathbf{E}_n+\mathbf{E}_n^*)$, due to linearity, it is sufficient to consider a fictitious single mode incident  field 
$\mathbf{E}_{inc} = \mathbf{E}_n + \mathbf{E}_n^*$ which impinges upon a closed PEC surface. The induced surface current $\mathbf{J}_s$ due to this incident field is determined by
\begin{equation} \label{EFIE-a}
\begin{aligned}
\mathcal{Z}_E(\mathbf{r},\mathbf{r}') \cdot \mathbf {J}_s(\mathbf{r}') & = - \hat n \times \left\lbrack\mathbf{E}_n(\mathbf{r}) + \mathbf{E}_n^*(\mathbf{r})\right\rbrack \\
& = -  \hat n \times \left\lbrack \mathcal{L}_E(\mathbf{r},\mathbf{r}') +\mathcal{L}_E^*(\mathbf{r},\mathbf{r}')  \right\rbrack \cdot \mathbf{J}_n(\mathbf{r}')\\
&= -2 \mathcal{R}_E (\mathbf{r},\mathbf{r}') \cdot \mathbf{J}_n(\mathbf{r}'), \quad \mathbf{r} \in S
\end{aligned}
\end{equation}
Due to Equation (\ref{E_GEP2}), it is obvious that
\begin{equation}\label{Js}
\mathbf{J}_s = \frac{-2}{1+i\lambda_n} \mathbf{J}_n
\end{equation}

One can also investigate the same problem with MFIE. It is easy to argue that the outgoing $\mathbf{E}_n$ corresponds to $\mathbf{H}_n$ and the incoming $\mathbf{E}_n^*$ corresponds to $-\mathbf{H}_n^*$ by using
\begin{equation}
\mathbf{H} = \frac{-i}{ k \eta}\nabla \times \mathbf{E}
\end{equation}
Therefore, the corresponding incident magnetic field is given by 
$\mathbf{H}_{inc} = \mathbf{H}_n - \mathbf{H}_n^*$, which is also source-free. The induced surface current $\mathbf{J}_s$ is determined by 
\begin{equation} \label{MFIE-a}
\begin{aligned}
\mathcal{Z}_H(\mathbf{r},\mathbf{r}') \cdot \mathbf{J}_s(\mathbf{r}') &= \overline{\mathbf{I}}_t \cdot \mathbf{H}_{inc}(\mathbf{r}) \\
& =  \overline{\mathbf{I}}_t \cdot \left\lbrack \mathbf{H}_n(\mathbf{r}) - \mathbf{H}_n^*(\mathbf{r}) \right \rbrack
, \quad \mathbf{r} \in S^+
\end{aligned}
\end{equation}
where $S^+$ represents a surface which is infinitesimally larger than $S$, $\overline{\mathbf{I}}_t = -\hat n \times \hat n \times = \overline{\mathbf{I}} - \hat n\hat n$ extracts the tangential field components, and
\begin{equation} 
\begin{aligned}
\mathcal{Z}_H(\mathbf{r},\mathbf{r}') \cdot \mathbf{J}(\mathbf{r}') & = \mathbf{J}(\mathbf{r}) \times \hat n - \overline{\mathbf{I}}_t \cdot \mathcal{K}_H(\mathbf{r},\mathbf{r}') \cdot \mathbf{J}(\mathbf{r}') 
\end{aligned}
\end{equation}
with
\begin{equation}
\begin{aligned}
\mathcal{K}_H(\mathbf{r},\mathbf{r}') \cdot \mathbf{J}(\mathbf{r}') &= \nabla \times \int_S d\mathbf{r}' \overline{\mathbf{G}}(\mathbf{r},\mathbf{r}') \cdot \mathbf{J}(\mathbf{r}') \\
& =  \int_S d \mathbf{r}' \nabla g(\mathbf{r},\mathbf{r}') \times \mathbf{J}(\mathbf{r}')
\end{aligned}
\end{equation}
When $\mathbf{r}\in S$, one has
\begin{equation} 
\begin{aligned}
\mathcal{Z}_H(\mathbf{r},\mathbf{r}') \cdot \mathbf{J}(\mathbf{r}') &= \frac{\mathbf{J}(\mathbf{r})}{2} \times \hat n \\
& \hspace{0.15in} - \overline{\mathbf{I}}_t \cdot P.V. \int_S d \mathbf{r}' \nabla g(\mathbf{r},\mathbf{r}') \times \mathbf{J}(\mathbf{r}')
\end{aligned}
\end{equation}
Note that $\mathbf{H}_n(\mathbf{r}) = \mathcal{K}_H(\mathbf{r},\mathbf{r}') \cdot \mathbf{J}_n(\mathbf{r}')$, Equation (\ref{MFIE-a}) gives rise to
\begin{equation}\label{MFIE-b}
\mathcal{Z}_H(\mathbf{r},\mathbf{r}') \cdot \mathbf{J}_s(\mathbf{r}') = -2i \mathcal{X}_{H}(\mathbf{r},\mathbf{r}') \cdot \mathbf{J}_n(\mathbf{r}') 
\end{equation}
where 
\begin{equation}
\mathcal{X}_H = \frac{1}{2i}\left(\mathcal{Z}_H - \mathcal{Z}_H^*\right)
\end{equation}
Since (\ref{MFIE-a}) and (\ref{EFIE-a}) govern the same physical problem, the same surface currents are induced. Substituting (\ref{Js}) into (\ref{MFIE-b}), the MFIE based TCM is obtained as
\begin{equation}\label{H-GEP2}
\mathcal{Z}_H(\mathbf{r},\mathbf{r}') \cdot \mathbf{J}_n (\mathbf{r}') = i (1+i \lambda_n) \mathcal{X}_{H}(\mathbf{r},\mathbf{r}') \cdot \mathbf{J}_n(\mathbf{r}'),\; \mathbf{r} \in S
\end{equation}
An equivalent form is given by
\begin{equation}\label{H-GEP1}
\mathcal{R}_H(\mathbf{r},\mathbf{r}') \cdot \mathbf{J}_n (\mathbf{r}') = - \lambda_n \mathcal{X}_{H}(\mathbf{r},\mathbf{r}') \cdot \mathbf{J}_n(\mathbf{r}'),\quad \mathbf{r} \in S
\end{equation}
where 
\begin{equation}
\mathcal{R}_H = \frac{1}{2}\left(\mathcal{Z}_H + \mathcal{Z}_H^*\right)
\end{equation}
Normally, EFIE and MFIE share a common set of characteristic pairs (currents and values) except the null space modes. Note that internal resonance modes of EFIE hardly radiate, while those of MFIE are induced currents which radiate efficiently. Hence, they do not enter the common set \cite{ChewBook2009}.   

To obtain discretized eigenvalue problems, RWG and $\hat n \times$RWG basis functions are chosen as $\mathbf{f}_j$ and $\mathbf{t}_i$, respectively. To achieve better accuracy, one can use Buffa-Christiansen (BC) or Chen-Wilton (CW) basis functions as expansion functions $\mathbf{t}_i$. Hence, (\ref{H-GEP1}) and (\ref{H-GEP2}) lead to
\begin{equation}\label{H_DGEP1}
\bar{\mathbf{R}}_H  \mathbf{J}_n = -\lambda_n \bar{\mathbf{X}}_H  \mathbf{J}_n 
\end{equation}
and
\begin{equation}\label{H_DGEP2}
\bar{\mathbf{Z}}_H  \mathbf{J}_n = i(1+i\lambda_n) \bar{\mathbf{X}}_H  \mathbf{J}_n 
\end{equation}
where $\bar{\mathbf{R}}_H$ and $\bar{\mathbf{X}}_H$ are the real and imaginary parts of $\bar{\mathbf{Z}}_H$, respectively, while $\bar{\mathbf{Z}}_H$ is the matrix representation of operator $\mathcal{Z}_H$.
Both $\bar{\mathbf{R}}_H$ and $\bar{\mathbf{Z}}_H$ become ill-conditioned near frequencies of internal resonances.   

\section{CFIE Based TCM}
Following the construction of CFIE for radiation or scattering problems \cite{Mautz78}, the CFIE based TCM is formulated as 
\begin{equation}
\begin{aligned}
&\left\lbrack \alpha \mathcal{Z}_E(\mathbf{r},\mathbf{r}') + (1-\alpha) \eta \mathcal{Z}_H(\mathbf{r},\mathbf{r}') \right \rbrack \cdot \mathbf{J}_n (\mathbf{r}') \\
&\quad= (1+i\lambda_n) \left\lbrack \alpha \mathcal{R}_E(\mathbf{r},\mathbf{r}') + (1-\alpha) i \eta
  \mathcal{X}_{H}(\mathbf{r},\mathbf{r}') \right\rbrack \cdot \mathbf{J}_n(\mathbf{r}')
\end{aligned}
\end{equation}
where $\mathbf{r} \in S$, and $\alpha$ is the combination coefficient. After discretization, the matrix eigenvalue equation is written as 
\begin{equation}\label{CFIE-TCM}
\bar{\mathbf{Z}}_C  \mathbf{J}_n = (1+i\lambda_n) \bar{\mathbf{K}}_C  \mathbf{J}_n
\end{equation}
where
\begin{subequations}
\begin{align}
\bar{\mathbf{Z}}_C &= \alpha \bar{\mathbf{Z}}_E + (1-\alpha) \eta \bar{\mathbf{Z}}_H \\
\bar{\mathbf{K}}_C & = \alpha \bar{\mathbf{R}}_E + (1-\alpha) i \eta \bar{\mathbf{X}}_{H} 
\end{align}
\end{subequations} 
Since $\bar{\mathbf{Z}}_C$ is full-rank even at frequencies of internal resonances, (\ref{CFIE-TCM}) is transformed to a standard eigenvalue equation as
\begin{equation}
\bar{\mathbf{Z}}_C^{-1} \bar{\mathbf{K}}_C  \mathbf{J}_n = (1+i\lambda_n)^{-1} \mathbf{J}_n
\end{equation} 
which can be iteratively solved.

Characteristic mode expansion of excited currents is formulated as follows. Consider an excitation problem which is determined by the CFIE as
\begin{equation}\label{CFIE}
\begin{aligned}
&\left\lbrack \alpha \mathcal{Z}_E(\mathbf{r},\mathbf{r}') + (1-\alpha) \eta \mathcal{Z}_H(\mathbf{r},\mathbf{r}') \right \rbrack \cdot \mathbf{J} (\mathbf{r}') \\
&\hspace{0.6in}=  -\alpha \hat n \times \mathbf{E}_{inc}(\mathbf{r}) + (1-\alpha) \eta
  \overline{\mathbf{I}}_t \cdot \mathbf{H}_{inc}(\mathbf{r})
\end{aligned}
\end{equation}
where $\mathbf{r} \in S$.
The discretization of (\ref{CFIE}) gives rise to 
\begin{equation}
\bar{\mathbf{Z}}_C  \mathbf{I}  = \mathbf{F}_{inc}
\end{equation}
where $\mathbf{I}$ is a column vector containing the expansion coefficients of RWGs $\mathbf{f}_j$, and $\mathbf{F}_{inc}$ is the vector representation of the mixed fields $-\alpha \hat n \times \mathbf{E}_{inc} + (1-\alpha) \eta \overline{\mathbf{I}}_t \cdot \mathbf{H}_{inc}$.  

Normally, $\bar{\mathbf{Z}}_C$ is not a symmetric matrix. One can construct an auxiliary eigenvalue problem of (\ref{CFIE-TCM}) as
\begin{equation}
\bar{\mathbf{Z}}_C^T \mathbf{J}_n^a = (1+i\lambda_n) \bar{\mathbf{K}}_C^T  \mathbf{J}_n^a
\end{equation}
where the superscript $T$ denotes the transpose of a matrix. The $\bar{\mathbf{K}}_C$-orthogonality is therefore obtained as
\begin{equation}
\left(\bar{\mathbf{J}}^{a}\right)^T \bar{\mathbf{K}}_C \bar{\mathbf{J}} = \bar{\mathbf{I}}
\end{equation} 
where matrices $\bar{\mathbf{J}}^{a}$ and $\bar{\mathbf{J}}$ contain characteristic currents $\mathbf{J}^a_n$ and $\mathbf{J}_n$ as columns, respectively, and $\bar{\mathbf{I}}$ is the identity matrix.
We assume $\mathbf{I}$ to be a linear superposition of $\mathbf{J}_n$, that is 
\begin{equation}\label{I=Ja}
\mathbf{I} =\bar{\mathbf{J}} \mathbf{a}
\end{equation}
where $\mathbf{a}$ is a vector containing the modal coefficients as entries.
Using the orthogonality condition, then
\begin{equation}
\left(\bar{\mathbf{J}}^{a}\right)^T \bar{\mathbf{Z}}_C \bar{\mathbf{J}} = \bar{\mathbf{\Sigma}}
\end{equation}
where the diagonal matrix $\bar{\mathbf{\Sigma}} = \textrm{diag}[1+i\lambda_1,1+i\lambda_2,\cdots]$, one can obtain the modal coefficients as 
\begin{equation}\label{cm-exp-coef}
 \mathbf{a} = \left(\bar{\mathbf{\Sigma}}\right)^{-1} \left(\bar{\mathbf{J}}^{a}\right)^T \mathbf{F}_{inc}
\end{equation}
Taking $\alpha=1$, the above is reduced to the conventional EFIE case where $\bar{\mathbf{Z}}_C$ becomes symmetric. 

\section{CMP-CFIE Based TCM}
In scattering or radiation problems, a Calder\'{o}n multiplicative preconditioner (CMP) has been used to accelerate the convergence of CFIE solutions \cite{Contopanagos02,Bagci09}. Similarly, we can obtain the CMP-CFIE based TCM as 
\begin{equation}
\begin{aligned}
&\left\lbrack \frac{\alpha}{\eta} \mathcal{Z}_E(ik) \cdot \mathcal{Z}_E(k) + (1-\alpha) \eta \mathcal{Z}_H(k) \right \rbrack  \mathbf{J}_n \\
&\quad= (1+i\lambda_n) \left\lbrack \frac{\alpha}{\eta} \mathcal{Z}_E(ik) \cdot \mathcal{R}_E(k) + (1-\alpha) i \eta
  \mathcal{X}_{H}(k) \right\rbrack  \mathbf{J}_n
\end{aligned}
\end{equation}
The discretization of the above simply follows \cite{Contopanagos02,Bagci09}, which leads to 
\begin{equation}
\bar{\mathbf{Z}}_{CC}  \mathbf{J}_n = (1+i\lambda_n) \bar{\mathbf{K}}_{CC}  \mathbf{J}_n
\end{equation} 
where 
\begin{subequations}
\begin{align}
\bar{\mathbf{Z}}_{CC} &= \frac{\alpha}{\eta} \tilde{\mathbf{Z}}_E \bar{\mathbf{G}}^{-1}  \bar{\mathbf{Z}}_E  + (1-\alpha) \eta \bar{\mathbf{Z}}_H \\
\bar{\mathbf{K}}_{CC} & = \frac{\alpha}{\eta} \tilde{\mathbf{Z}}_E \bar{\mathbf{G}}^{-1} \bar{\mathbf{R}}_E + (1-\alpha) i \eta \bar{\mathbf{X}}_{H} 
\end{align}
\end{subequations} 
In the above, $\tilde{\mathbf{Z}}_E$ is the matrix representation of operator $\mathcal{Z}_E(ik)$ with BCs and $\hat n\times$BCs chosen as the expansion and testing functions, respectively, and $\bar{\mathbf{G}}$ is the Gramian matrix linking $\hat n\times$RWGs and BCs \cite{Andriulli}. It has been shown that $\bar{\mathbf{Z}}_{CC}$ is well-conditioned even at frequencies of internal resonances.


\section{Numerical Results} 
In this study, the discretized standard eigenvalue problems are solved with the popular IRAM embedded in MATLAB R2014b on an Intel Core i7-4700MQ CPU with $2.40$ GHz clock rate. In each IRAM iteration, we compute the MVM $\bar{\mathbf{Z}}^{-1} \mathbf{u}$ with the generalized minimal residual (GMRES) method, where $\bar{\mathbf{Z}}$ can be $\bar{\mathbf{Z}}_E$, $\bar{\mathbf{Z}}_H$, $\bar{\mathbf{Z}}_{C}$,  $\bar{\mathbf{Z}}_{CC}$, etc. In each GMRES iteration, an MVM in the form of $\bar{\mathbf{Z}} \mathbf{u}$ is directly computed for convenience, where $\bar{\mathbf{Z}}$ is explicitly stored. For large-scale applications, one can apply MLFMA to compute MVMs such as $\bar{\mathbf{Z}} \mathbf{u}$, $\bar{\mathbf{R}} \mathbf{u}$ and $\bar{\mathbf{X}} \mathbf{u}$  implicitly with a complexity of $O(N\log N)$ \cite{Dai14}. In the following examples, the total number of IRAM (outer) iterations is denoted as $N_{out}$, and the average number of GMRES (inner) iterations in each IRAM (outer) iteration is denoted as $N_{in}$. The GMRES error tolerance is set to $10^{-10}$ unless specified otherwise. 

\begin{figure}[]
\centering
\includegraphics[scale=0.5]{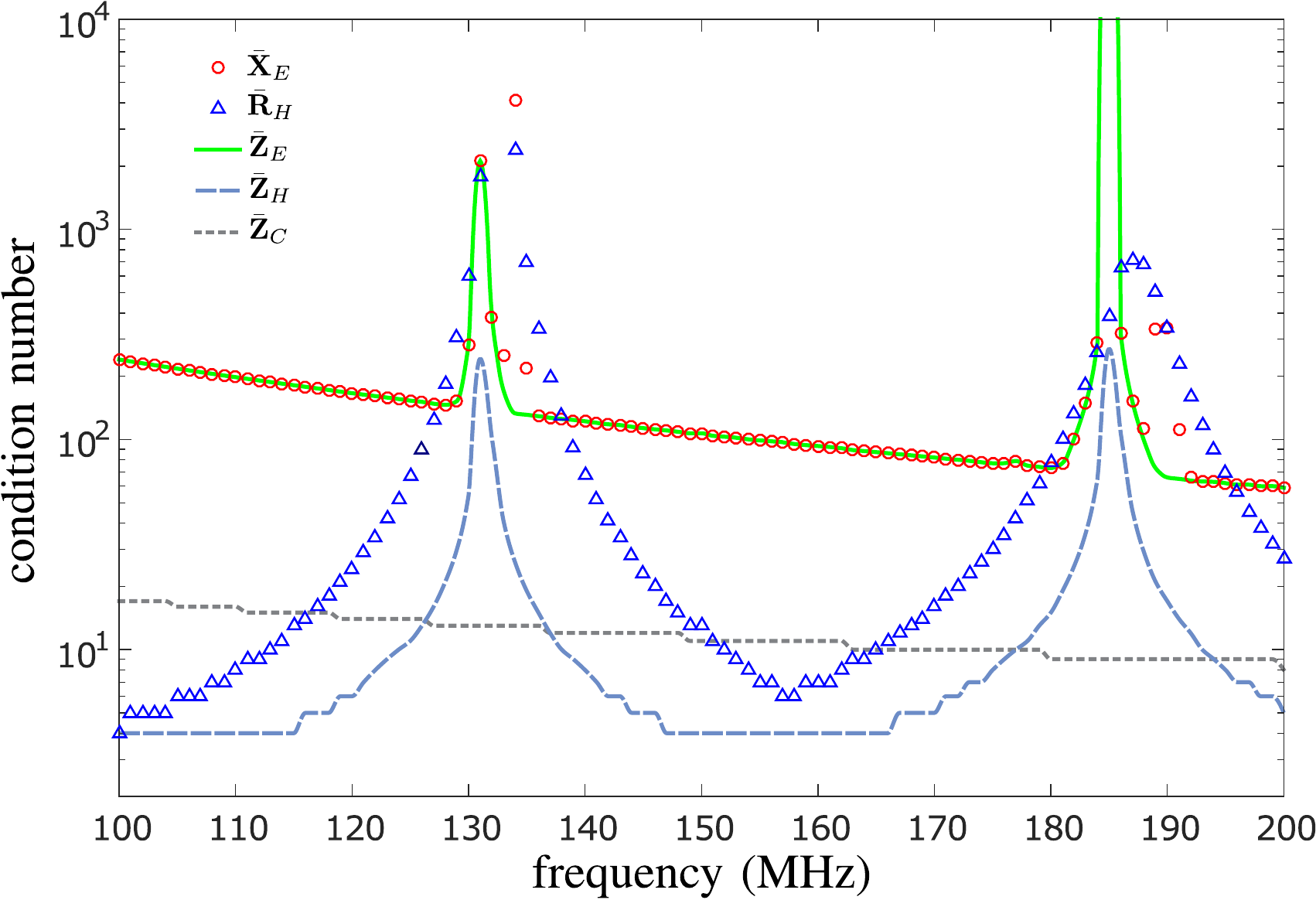}
\caption{Condition numbers of EFIE, MFIE and CFIE matrices. It clearly indicates that CFIE removes the corruption due to internal resonance modes.} \label{sphSweepCond}
\end{figure}

We first consider a perfectly conducting sphere with a radius of $1$~m where $2\, 280$ unknowns are used. The condition numbers of matrices $\bar{\mathbf{Z}}_E$, $\bar{\mathbf{X}}_E$, $\bar{\mathbf{Z}}_H$, $\bar{\mathbf{R}}_H$ and $\bar{\mathbf{Z}}_C$ are plotted in Fig.~\ref{sphSweepCond}, where the frequency is swept at a step of $1$~MHz. The spikes in the condition numbers of $\bar{\mathbf{Z}}_E$ correspond to spurious internal resonances of EFIE. It is obvious that $\bar{\mathbf{Z}}_E$ and $\bar{\mathbf{X}}_E$ have the same condition numbers at most frequencies except that $\bar{\mathbf{X}}_E$ seems to have multiple spikes around each internal resonance frequency. This renders it difficult to compute characteristic modes using  (\ref{E_DSEP1}) and (\ref{E_DSEP2}).  Although MFIE is usually well-conditioned, it also suffers from internal resonances as several spikes are observed in the condition numbers of $\bar{\mathbf{Z}}_H$. So does $\bar{\mathbf{R}}_H$ except that it is more poorly-conditioned. Hence,  (\ref{H_DGEP1}) and (\ref{H_DGEP2}) may not be suitable for characteristic mode computation when the operating frequency is near internal resonances. On the other hand, CFIE is free from the internal resonance corruption as no spikes are observed in the condition numbers of $\bar{\mathbf{Z}}_C$  within the entire frequency band. By setting the CFIE combination coefficient $\alpha$ to $0.5$, $\bar{\mathbf{Z}}_C$ has a small condition number ranging from $8$ to $17$.  

\begin{figure}[]
\centering
\subfigure[]{\includegraphics[scale=0.36]{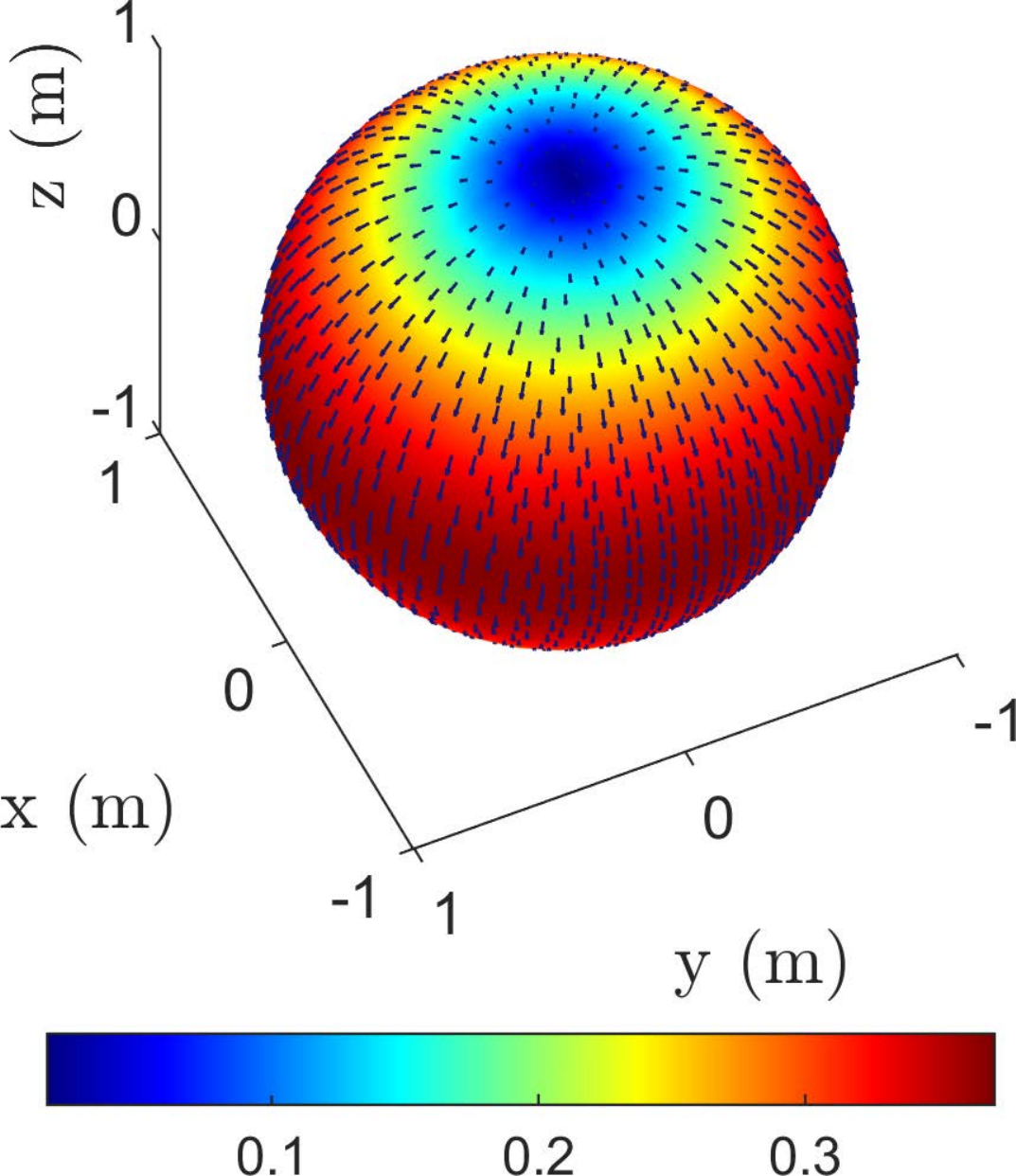} \label{sph-Js-131p192m-efie}}
\subfigure[]{\includegraphics[scale=0.36]{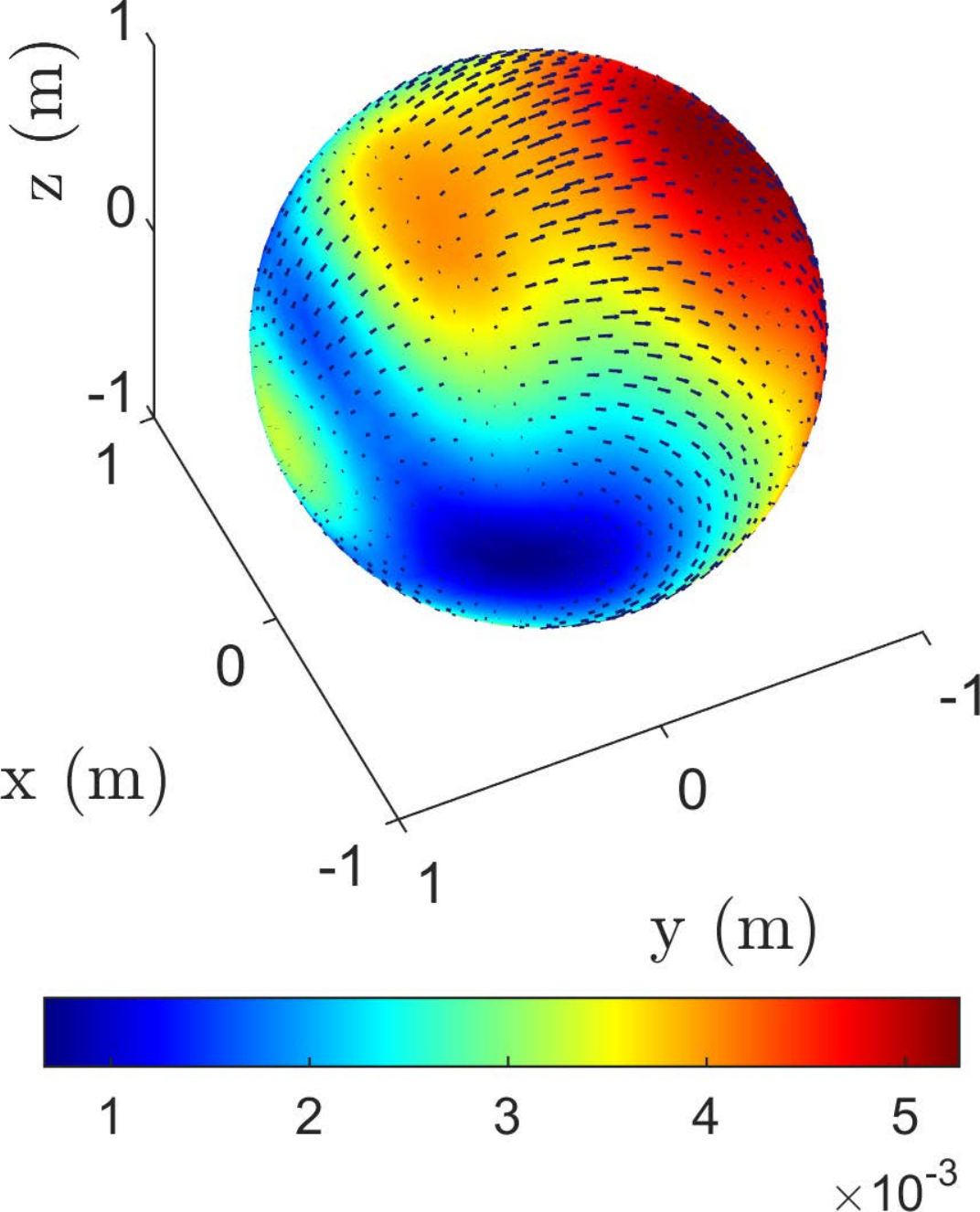} \label{sph-Js-131p192m-cfie}} 
\caption{Surface currents of a sphere excited by a plane wave at $131.192$~MHz: (a) Incorrect current computed by EFIE which is dominated by a spurious internal resonance mode. (b) Correct induced current computed by CFIE ($0.5$). } \label{sph-Js}
\end{figure}

A finer frequency sweep is performed with respect to the spikes of $\bar{\mathbf{Z}}_E$. Two lowest frequencies of internal resonances are located with better accuracies, which are listed in Table~\ref{tabsphCondNum}, with the condition numbers of $\bar{\mathbf{Z}}_E$, $\bar{\mathbf{X}}_E$, $\bar{\mathbf{Z}}_H$, $\bar{\mathbf{R}}_H$ and $\bar{\mathbf{Z}}_C$ provided. It clearly shows that $\bar{\mathbf{Z}}_C$ is immune to the internal resonance corruption which the other matrices are susceptible to. At the internal resonance frequency $131.192$~MHz, we excite the sphere by a $\hat z$-polarized plane wave propagating along  $\hat y$-direction. The induced current computed by EFIE, as shown in Fig.~\ref{sph-Js-131p192m-efie}, is incorrect as it is dominated by a spurious internal resonance mode. By solving a standard eigenvalue problem $\bar{\mathbf{Z}}_E \mathbf{u}_n = \zeta_n \mathbf{u}_n$, one can easily find the internal resonance mode which corresponds to the smallest $\left\vert \zeta_n \right\vert$ (around zero). The induced current can be correctly solved for by CFIE with a combination coefficient $\alpha=0.5$, as shown in Fig.~\ref{sph-Js-131p192m-cfie}.

We then compute $100$ modes with the smallest $|\lambda_n|$ by EFIE, MFIE and CIFE, and list the first few characteristic values $\lambda_n$ in Table~\ref{sphCV}. It is observed that the discrepancies between the results computed by EFIE and MFIE are larger for the first $3$ characteristic values, which correspond to MFIE internal resonance modes. Since MFIE is not implemented to generate so accurate solutions as EFIE, the numerically computed internal resonance frequencies of the two are not exactly the same, and CFIE leads to complex $\lambda_n$ with small imaginary parts. Even so, the real parts of $\lambda_n$ obtained by CFIE agree well with EFIE results. 

\begin{figure}[!t]
	\centering
	\subfigure[]{\includegraphics[scale=0.33]{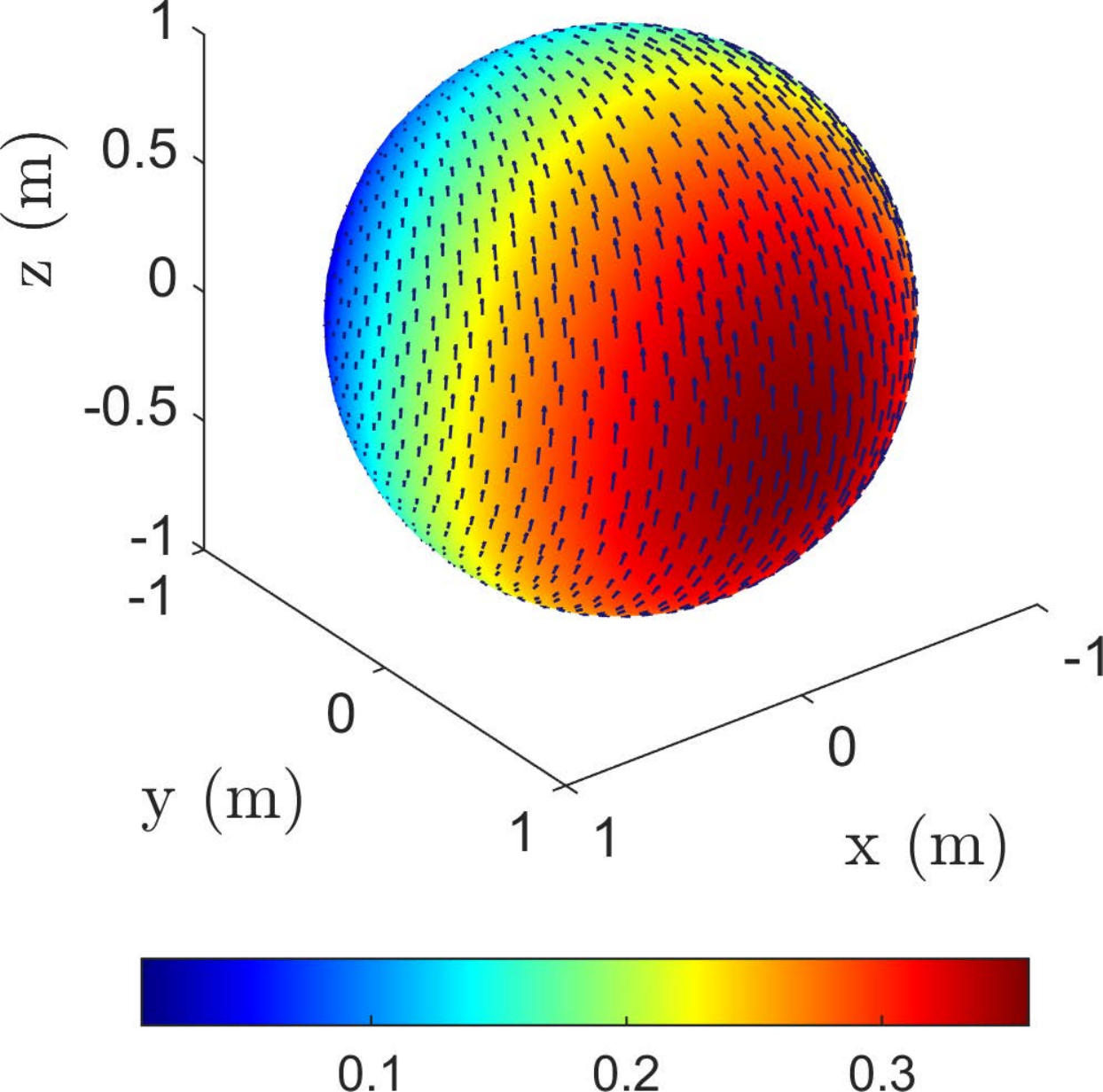} \label{sph-J3-131p192m-efie}}
	\subfigure[]{\includegraphics[scale=0.33]{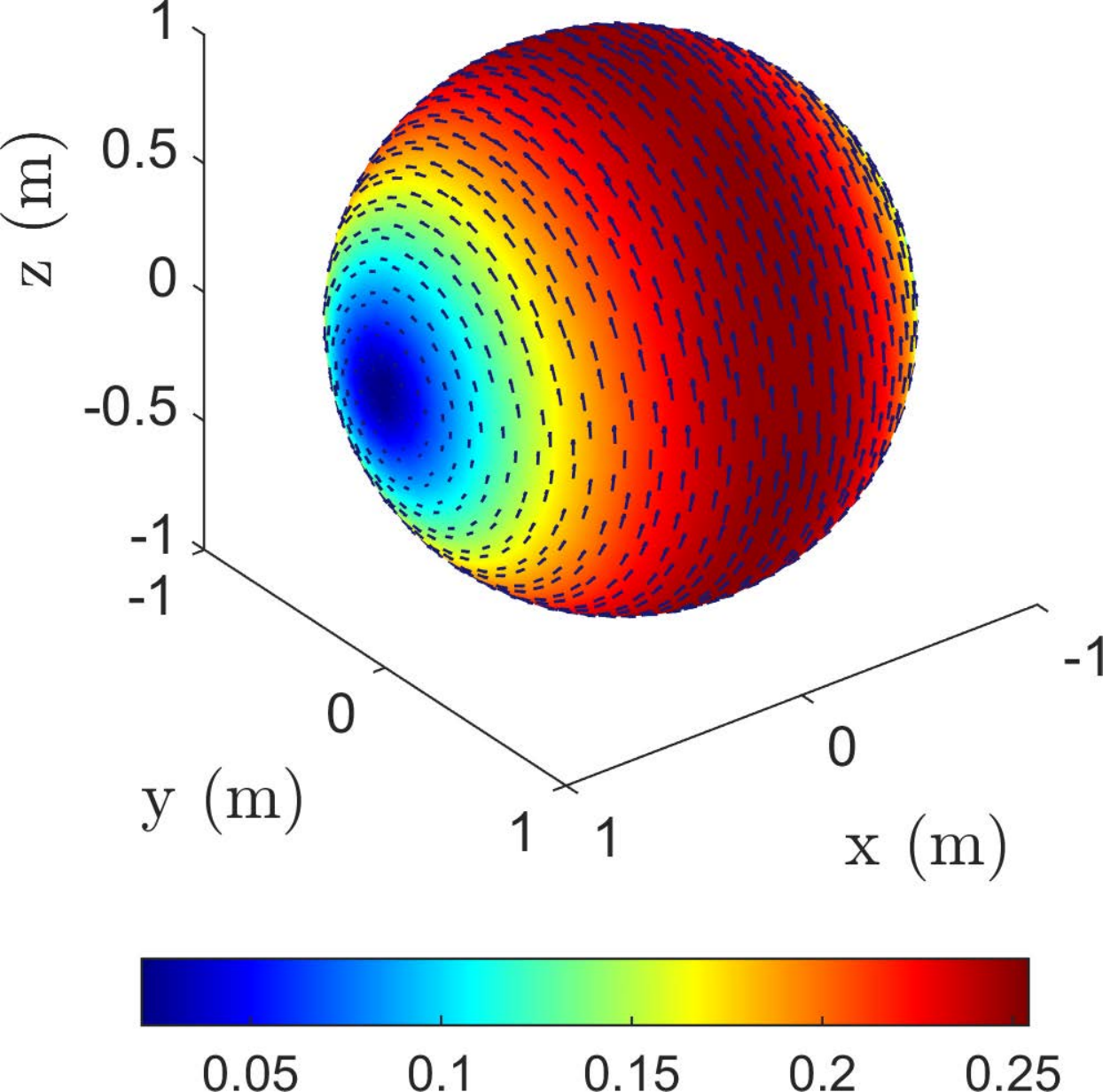} \label{sph-J3-131p192m-cfie}} 
		\subfigure[]{\includegraphics[scale=0.33]{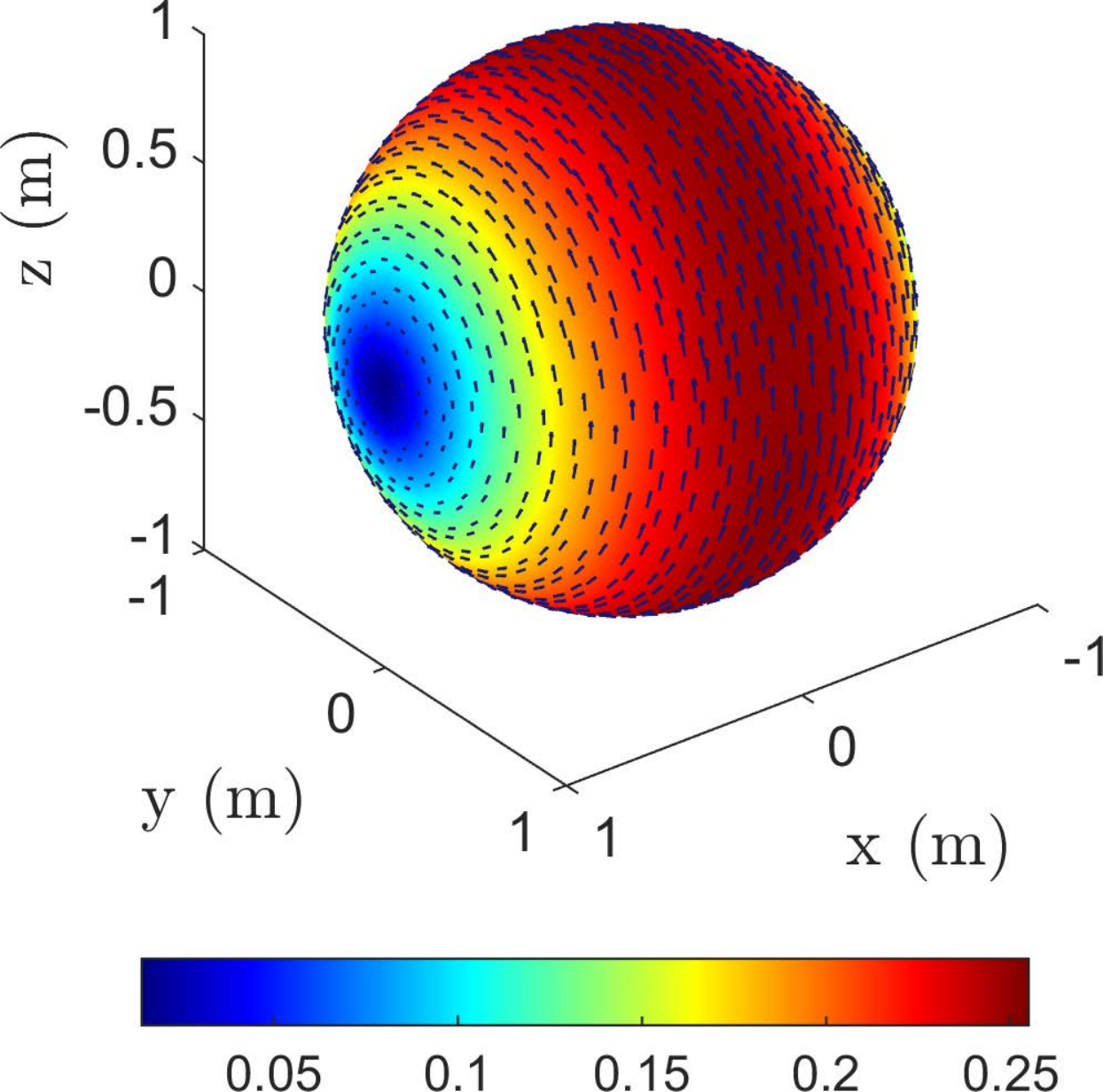} \label{sph-J3-128m-efie}}
		\subfigure[]{\includegraphics[scale=0.33]{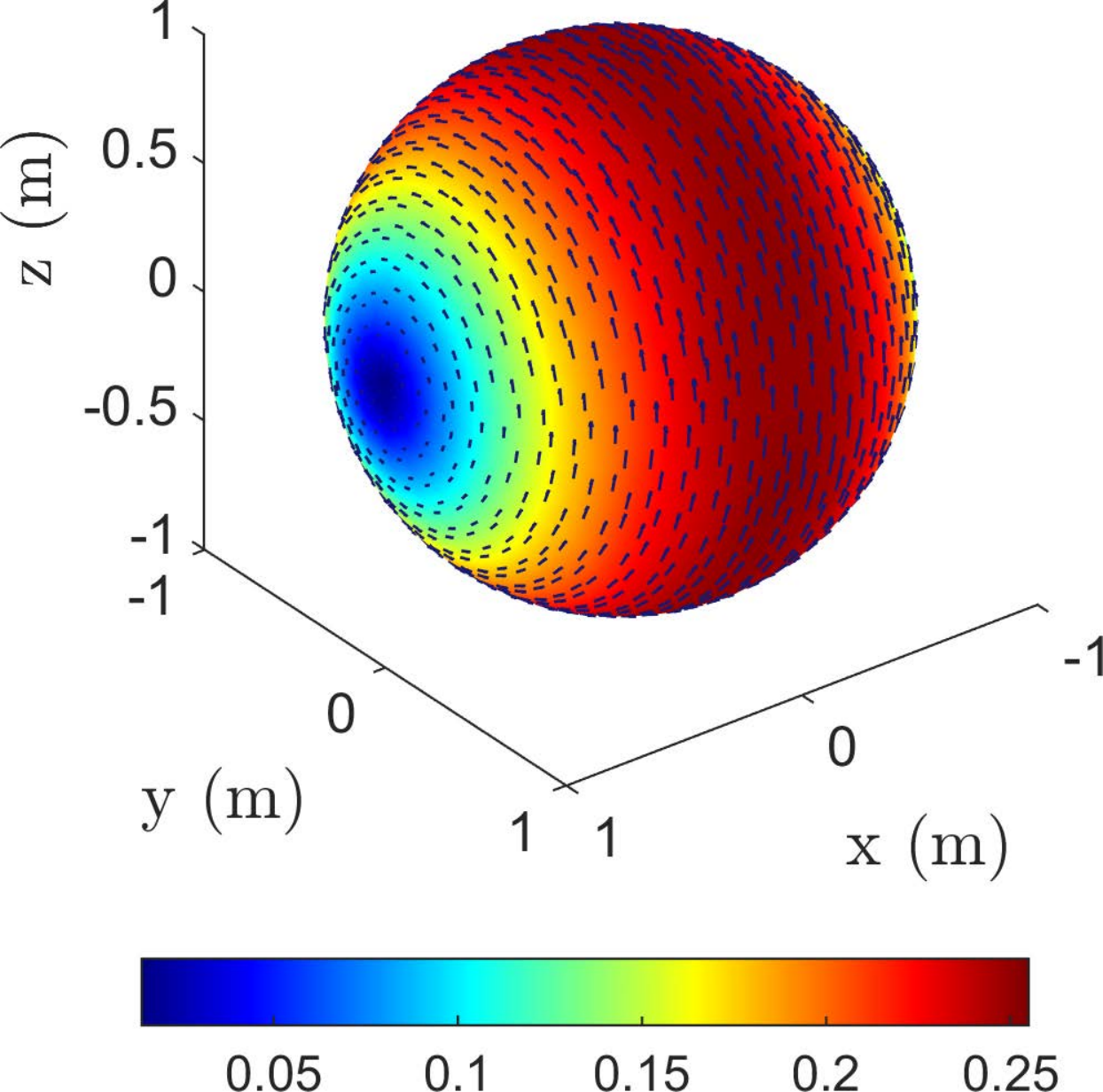} \label{sph-J1-136m-efie}} 
	\caption{Sphere modal current $\mathbf{J}_3$: (a) Pattern inconsistent with others, computed by EFIE at $131.192$~MHz. (b) Correct pattern computed by CFIE ($0.5$) at $131.192$~MHz. (c) Correct pattern computed by EFIE at $128$~MHz. (d) Correct pattern computed by EFIE at $136$~MHz. The corresponding characteristic values of the $4$ currents are $0.0509$, $0.0509-0.0004i$, $0.1100$, and $-0.0382$, respectively.} \label{sph-J3}
\end{figure}

\begin{figure}[]
	\centering
	\includegraphics[scale=0.6]{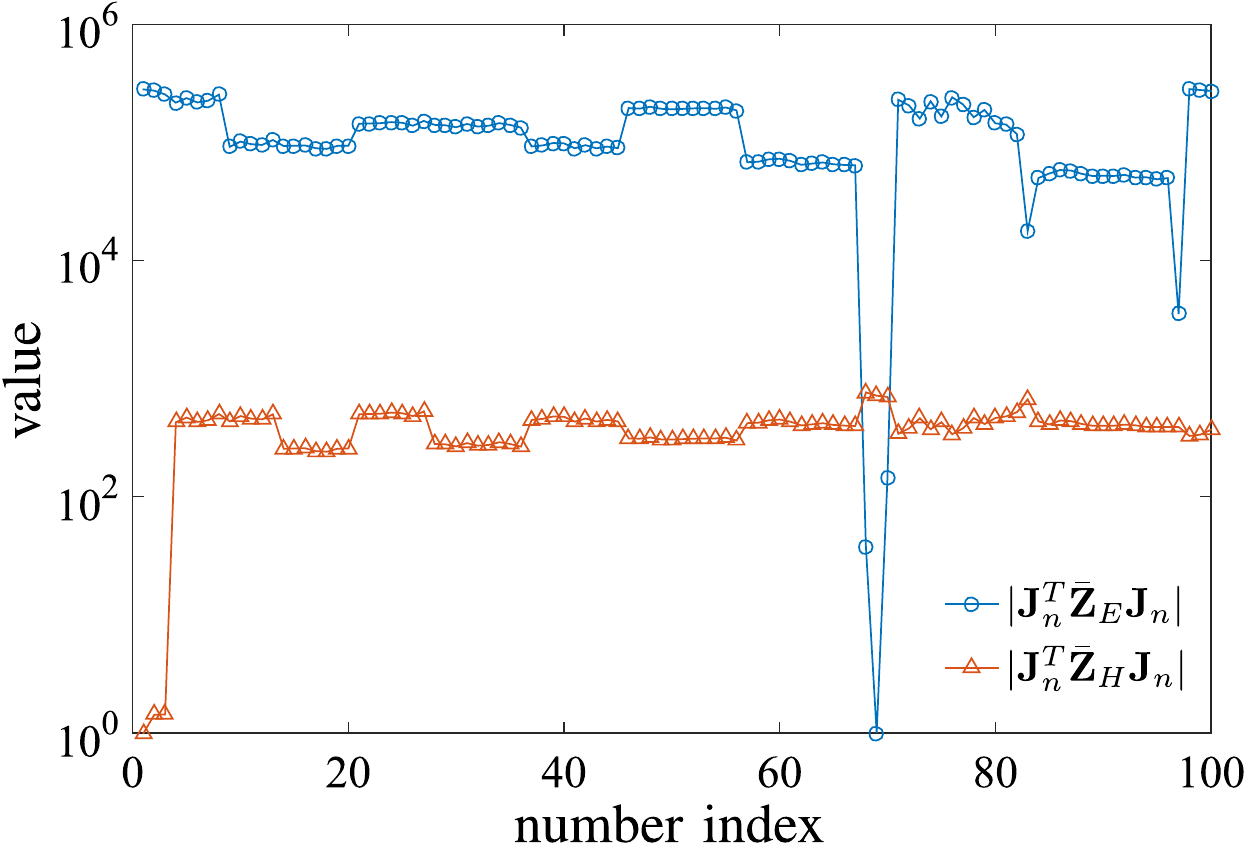}
	\caption{Normalized quantities $|\mathbf{J}^T_n \bar{\mathbf{Z}}_E \mathbf{J}_n|$ and $|\mathbf{J}^T_n \bar{\mathbf{Z}}_H \mathbf{J}_n|$ of the first $100$ modes where $\mathbf{J}_n$ are computed by CFIE at $131.192$~MHz.} \label{JZJ}
\end{figure}

At frequencies of spurious internal resonances, EFIE may not be able to find correct modal currents. For example, we plot in Fig.~\ref{sph-J3-131p192m-efie} the third modal current $\mathbf{J}_3$ computed by EFIE at $131.192$~MHz. It is not consistent with those found by EFIE at a lower frequency, e.g., $128$~MHz [Fig.~\ref{sph-J3-128m-efie}], or a higher frequency, e.g., $136$~MHz [Fig.~\ref{sph-J1-136m-efie}]. However, the correct modal current can be obtained with CFIE even at internal resonances, as shown in Fig.~\ref{sph-J3-131p192m-cfie}. Furthermore, in Fig.~\ref{JZJ}, we plot the normalized quantities $|\mathbf{J}^T_n \bar{\mathbf{Z}}_E \mathbf{J}_n|$ and $|\mathbf{J}^T_n \bar{\mathbf{Z}}_H \mathbf{J}_n|$ for the first $100$ modes, where $\mathbf{J}_n$ are computed by CFIE at $131.192$~MHz. Obviously, the first $3$ modes correspond to null space modes of $\bar{\mathbf{Z}}_H$ which are well radiating. Moreover, $\mathbf{J}_{68}$ ($\lambda_{68}=-739$), $\mathbf{J}_{69}$ ($\lambda_{69}=-864$) and $\mathbf{J}_{70}$ ($\lambda_{70}=-2731$) are null space modes of $\bar{\mathbf{Z}}_E$ which are poorly radiating. They have similar current patterns as shown in Fig.~\ref{sph-Js-131p192m-efie}. Table~\ref{sphIterNum} shows $N_{out}$, $N_{in}$ and CPU times of different TCMs, where EFIE and EFIE$^\star$ denote the cases of using (\ref{E_DSEP2}) and (\ref{E_DSEP1}), respectively. It is obvious that CFIE based TCM is the most efficient, even though EFIE$^\star$ involves real number evaluation only. Besides, in Table~\ref{sphCV2}, we compare a few characteristic values computed by EFIE, MFIE and CFIE when the operating frequency is shifted to $128$~MHz. In this case, better consistency is observed between the results of EFIE (CFIE) and MFIE based TCMs.

\begin{table}[]
\renewcommand{\arraystretch}{1.3}
\caption{Condition Numbers of EFIE, MFIE, and CFIE Matrices at Internal Resonances (Sphere)}
\vskip0.in
\begin{center}
\setlength{\tabcolsep}{.6em}
\small \begin{tabular}{c c c c c c c}\hline
Freq (MHz) & $\bar{\mathbf{Z}}_E$ & $\bar{\mathbf{X}}_E$ & $\bar{\mathbf{Z}}_H$ & $\bar{\mathbf{R}}_H$ & $\bar{\mathbf{Z}}_{C} $  \\ \hline
$131.192$ & $1.02\times 10^6$ & $1.06\times 10^6$ & $839$ & $2\,437$ & $13$  \\
$185.005$ & $6.13\times 10^5$ & $6.44\times 10^5$ & $275$ & $388$ & $9$ \\ \hline
\end{tabular}
\end{center}
\label{tabsphCondNum}
\end{table}

\begin{table}[]
	\renewcommand{\arraystretch}{1.3}
	\caption{Characteristic Values (CVs) Computed by EFIE, MFIE, and CFIE at Internal Resonance $131.192$~MHz (Sphere)}
	\vskip0.in
	\begin{center}
		\setlength{\tabcolsep}{1.em}
		\small \begin{tabular}{c c c c}\hline
			CVs & EFIE & MFIE & CFIE \\ \hline
			$\lambda_1$ & $0.0503$ &$0.2457$  & $0.0503-0.0005i$  \\ 
			$\lambda_2$ & $0.0508$ & $0.2697$ & $0.0508-0.0004i$  \\
			$\lambda_3$ & $0.0509$ & $0.5293$ & $0.0509-0.0004i$  \\
			$\lambda_4$ & $1.2628$ & $1.2716$ & $1.2680-0.0026i$  \\ 
			$\lambda_5$ & $1.2660$ & $1.2720$ & $1.2680-0.0031i$  \\
			$\lambda_6$ & $1.2666$ & $1.2722$ & $1.2682-0.0026i$  \\
			$\lambda_7$ & $1.2670$ & $1.2727$ & $1.2683-0.0022i$  \\
			$\lambda_8$ & $1.2675$ & $1.2734$ & $1.2693-0.0027i$  \\
			$\lambda_9$ & $-1.3233$ & $-1.3293$ & $-1.3277-0.0026i$  \\
			$\lambda_{10}$ & $-1.3237$ & $-1.3299$ & $-1.3280-0.0029i$  \\ \hline
		\end{tabular}
	\end{center}
	\label{sphCV}
\end{table}

\begin{table}[]
	\renewcommand{\arraystretch}{1.3}
	\caption{Iteration Numbers and CPU Times of EFIE and CFIE at Internal Resonance $131.192$~MHz (Sphere)}
	\vskip0.in
	\begin{center}
		\setlength{\tabcolsep}{1.43em}
		\small \begin{tabular}{c c c c }\hline
			TCM & EFIE & EFIE$^\star$ & CFIE  \\ \hline
			$N_{out}$ & $100$ & $101$ & $100$ \\ 
			$N_{in}$ & $381.4$ & $391.2$ & $78.6$  \\
			CPU time (secs) & $789.9$ & $300.5$ & $99.9$  \\ \hline
		\end{tabular}
	\end{center}
	\label{sphIterNum}
\end{table}

\begin{table}[]
	\renewcommand{\arraystretch}{1.3}
	\caption{Characteristic Values (CVs) Computed by EFIE, MFIE, and CFIE at $128$~MHz (Sphere)}
	\vskip0.in
	\begin{center}
		\setlength{\tabcolsep}{1.em}
		\small \begin{tabular}{c c c c}\hline
			CVs & EFIE & MFIE & CFIE \\ \hline
			$\lambda_1$ & $0.1094$ &$0.1157$  & $0.1095-0.0004i$  \\ 
			$\lambda_2$ & $0.1099$ & $0.1159$ & $0.1099-0.0003i$  \\
			$\lambda_3$ & $0.1100$ & $0.1183$ & $0.1100-0.0003i$  \\
			$\lambda_4$ & $-1.2755$ & $-1.2814$ & $-1.2797-0.0026i$  \\ 
			$\lambda_5$ & $-1.2760$ & $-1.2819$ & $-1.2799-0.0029i$  \\
			$\lambda_6$ & $-1.2779$ & $-1.2837$ & $-1.2818-0.0027i$  \\
			$\lambda_7$ & $-1.2795$ & $-1.2851$ & $-1.2833-0.0026i$  \\
			$\lambda_8$ & $-1.2808$ & $-1.2861$ & $-1.2844-0.0025i$  \\
			$\lambda_9$ & $1.3824$ & $1.3881$ & $1.3846-0.0025i$  \\
			$\lambda_{10}$ & $1.3828$ & $1.3885$ & $1.3846-0.0032i$  \\ \hline
		\end{tabular}
	\end{center}
	\label{sphCV2}
\end{table}

\begin{figure}[!t]
	\centering
	\subfigure[]{\includegraphics[scale=0.28]{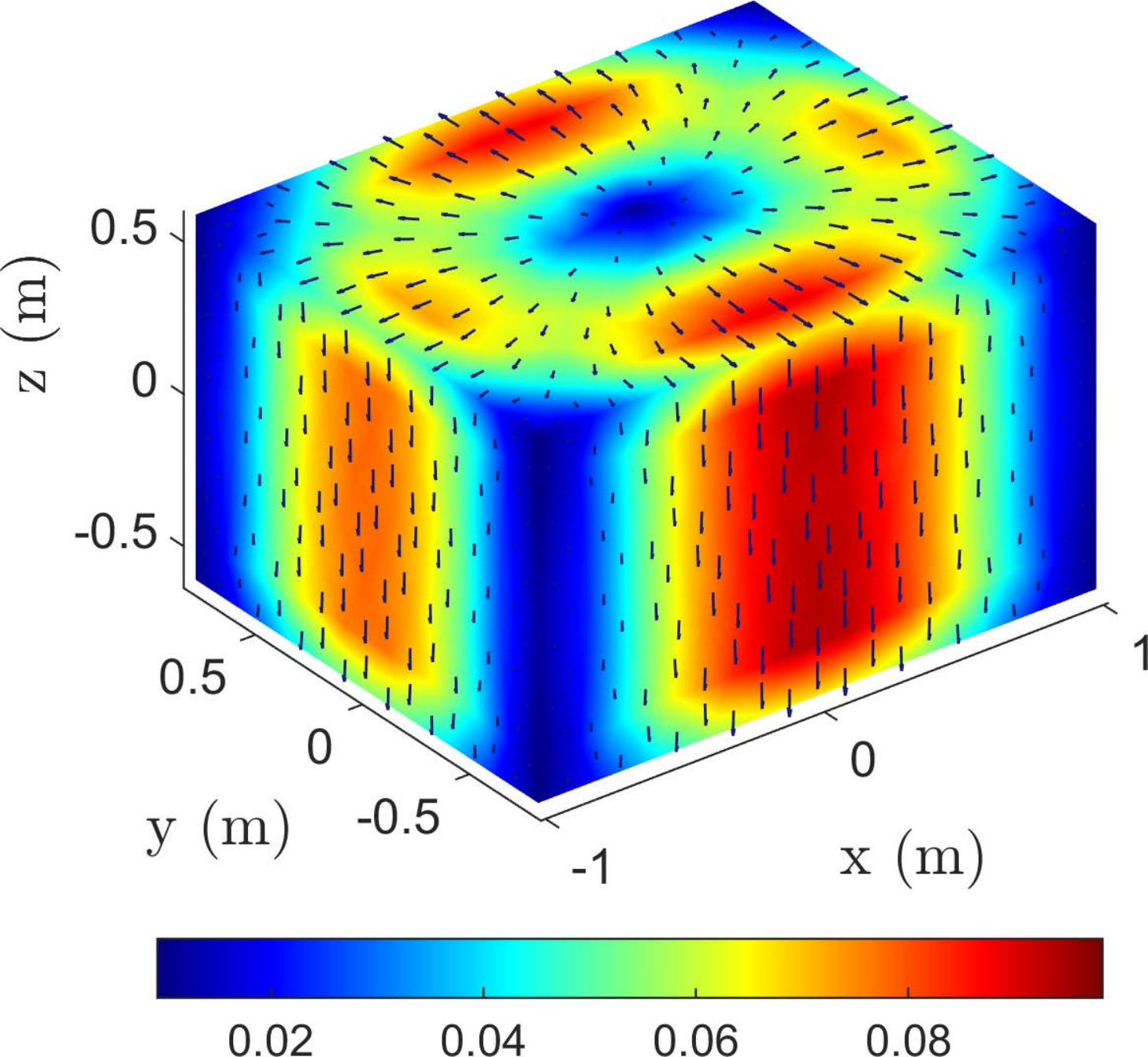} \label{cub-Js-119p88m-efie}}
	\subfigure[]{\includegraphics[scale=0.28]{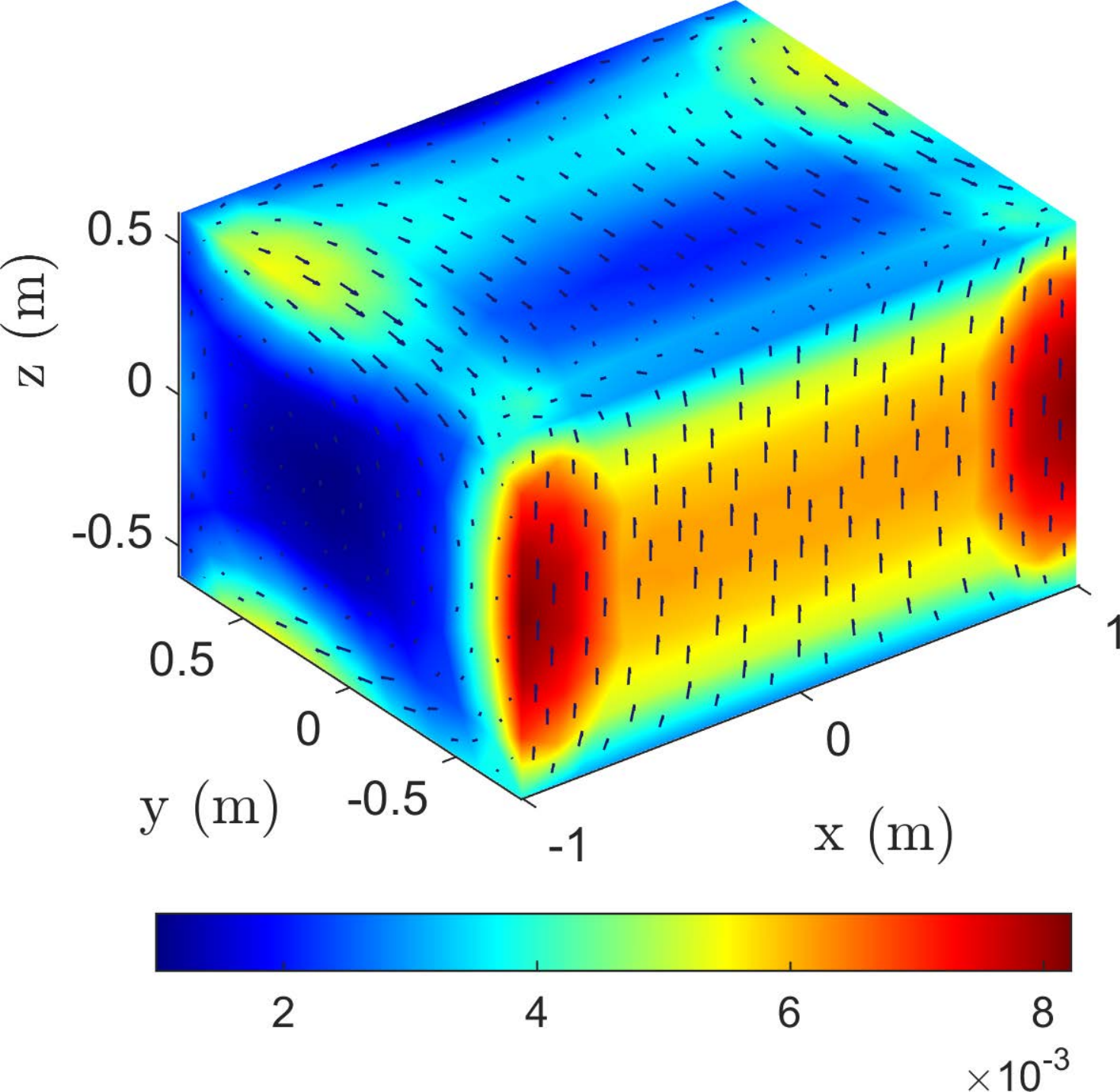} \label{cub-Js-119p98m-cmpcfie}} 
	\caption{Surface currents of a cuboid excited by a plane wave at $119.88$~MHz: (a) Incorrect induced current computed by EFIE which is dominated by a spurious internal resonance mode. (b) Correct induced current computed by CMP-CFIE ($0.9$).  } \label{cub-Js}
\end{figure}

\begin{figure}[!t]
	\centering
	\includegraphics[scale=0.5]{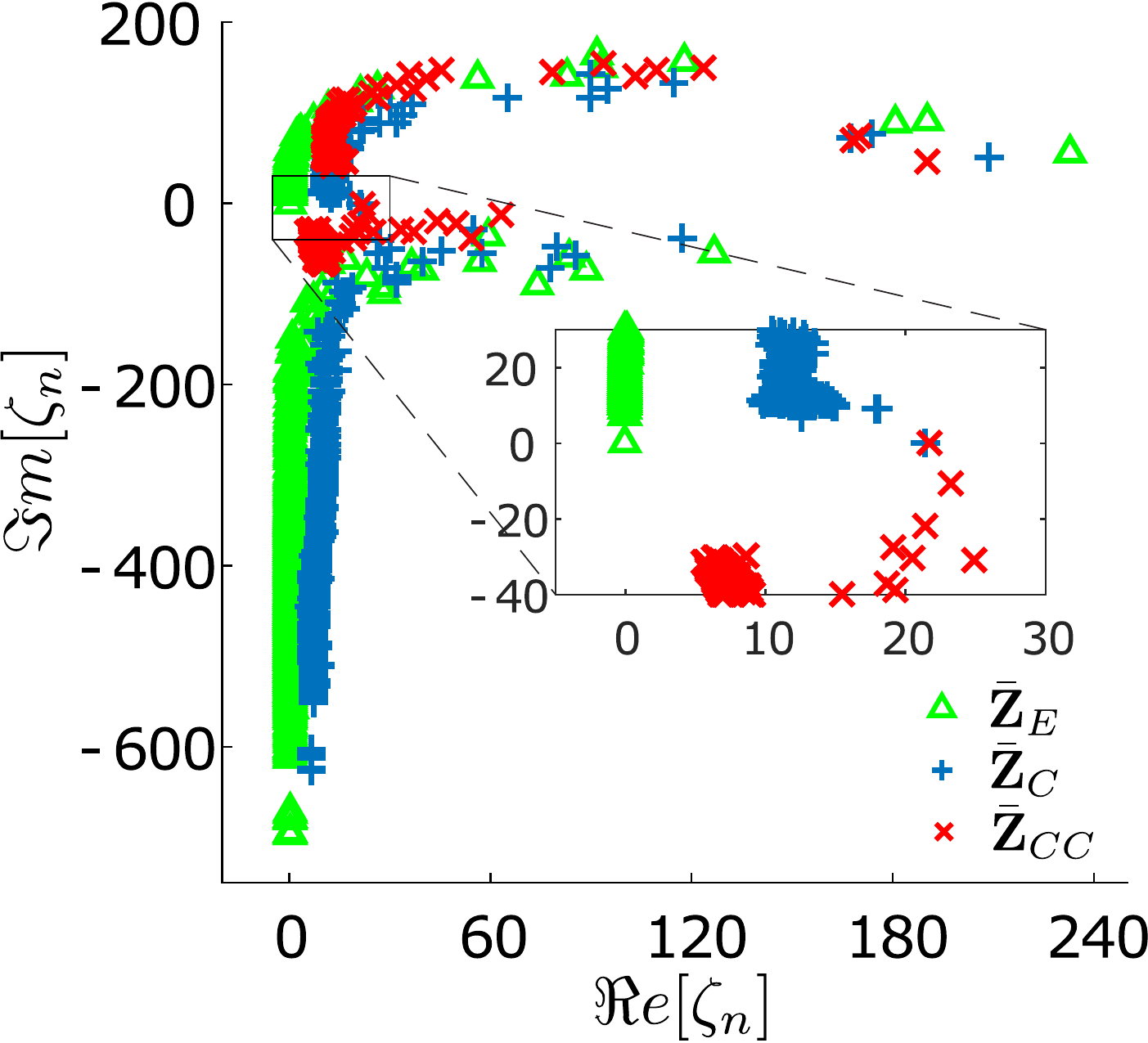}
	\caption{Eigenvalues $\zeta_n$ of matrices $\bar{\mathbf{Z}}_E$, $\bar{\mathbf{Z}}_C$ and $\bar{\mathbf{Z}}_{CC}$ at internal resonance. No zero eigenvalues are found for CFIE and CMP-CFIE. CMP-CFIE has the most compact eigen-spectral distribution.} \label{eigspectrum}
\end{figure}

We next consider a  cuboid with a dimension of $2.0 \times 1.6 \times 1.2$~m  where $1 \, 311$ edges are used. The lowest frequency of spurious internal resonances is found to be $119.88$~MHz using a fine frequency sweep. By setting the operating frequency to $119.880$~MHz, we excite the cuboid with a $\hat z$-polarized plane wave propagating along  $\hat y$-direction. The induced current computed by EFIE is incorrect as it is dominated by a spurious internal resonance mode [Fig.~\ref{cub-Js-119p88m-efie}], which again corresponds to a null space vector of $\bar{\mathbf{Z}}_E$. The correct induced current is computed by CMP-CFIE with a combination coefficient $\alpha=0.9$ as demonstrated in Fig.~\ref{cub-Js-119p98m-cmpcfie}. Table~{\ref{cubCondNum}} lists the condition numbers of $\bar{\mathbf{Z}}_E$, $\bar{\mathbf{Z}}_H$, $\bar{\mathbf{Z}}_C$, and $\bar{\mathbf{Z}}_{CC}$ at an internal resonance frequency, e.g., $119.88$~MHz, and a non-internal resonance frequency, e.g., $121$~MHz, respectively. It is obvious that the CFIE matrix $\bar{\mathbf{Z}}_C$ is well-conditioned at the internal resonance frequency, while the CMP-CFIE one $\bar{\mathbf{Z}}_{CC}$ is even better conditioned. This is reflected by the eigen-spectral distributions of $\bar{\mathbf{Z}}_E$, $\bar{\mathbf{Z}}_C$ and $\bar{\mathbf{Z}}_{CC}$, as plotted in Fig.~\ref{eigspectrum}. 


\begin{table}[!t]
\renewcommand{\arraystretch}{1.3}
\caption{Condition Numbers of EFIE, MFIE, CFIE and CMP-CFIE Matrices (Cuboid)}
\vskip0.in
\begin{center}
\setlength{\tabcolsep}{1.em}
\small \begin{tabular}{c c c c c}\hline
Freq (MHz) & $\bar{\mathbf{Z}}_E$ & $\bar{\mathbf{Z}}_H$ & $\bar{\mathbf{Z}}_C$ & $\bar{\mathbf{Z}}_{CC}$  \\ \hline
$119.88$ & $7.4305\times10^5$ & $106$ & $58$ & $10$ \\ 
$121$ & $276$ & $65$ & $44$ & $10$  \\\hline
\end{tabular}
\end{center}
\label{cubCondNum}
\end{table}

\begin{figure}[!t]
\centering
\subfigure[]{\includegraphics[scale=0.27]{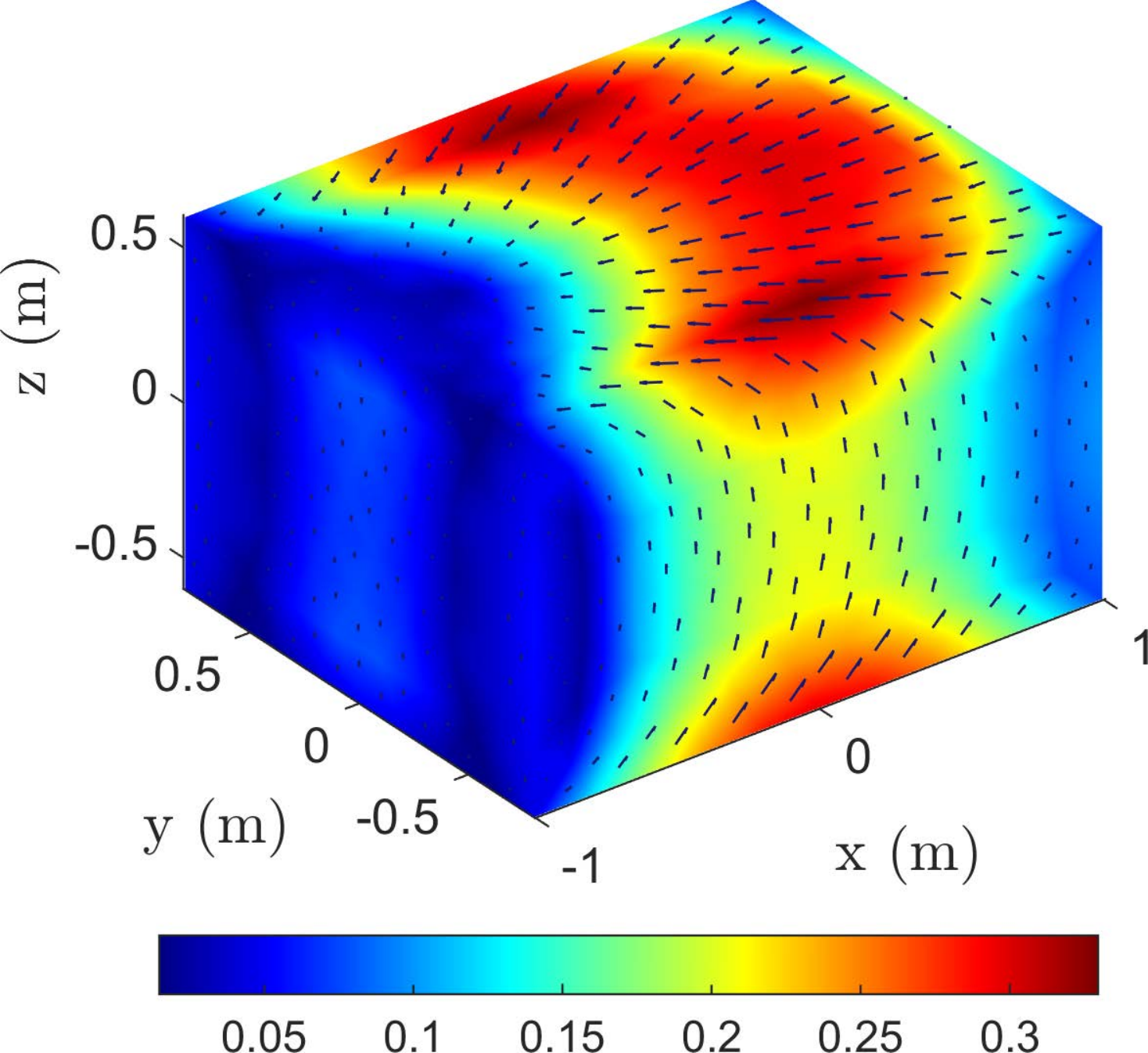} \label{cub-J3-119p88m-efie}} 
\subfigure[]{\includegraphics[scale=0.27]{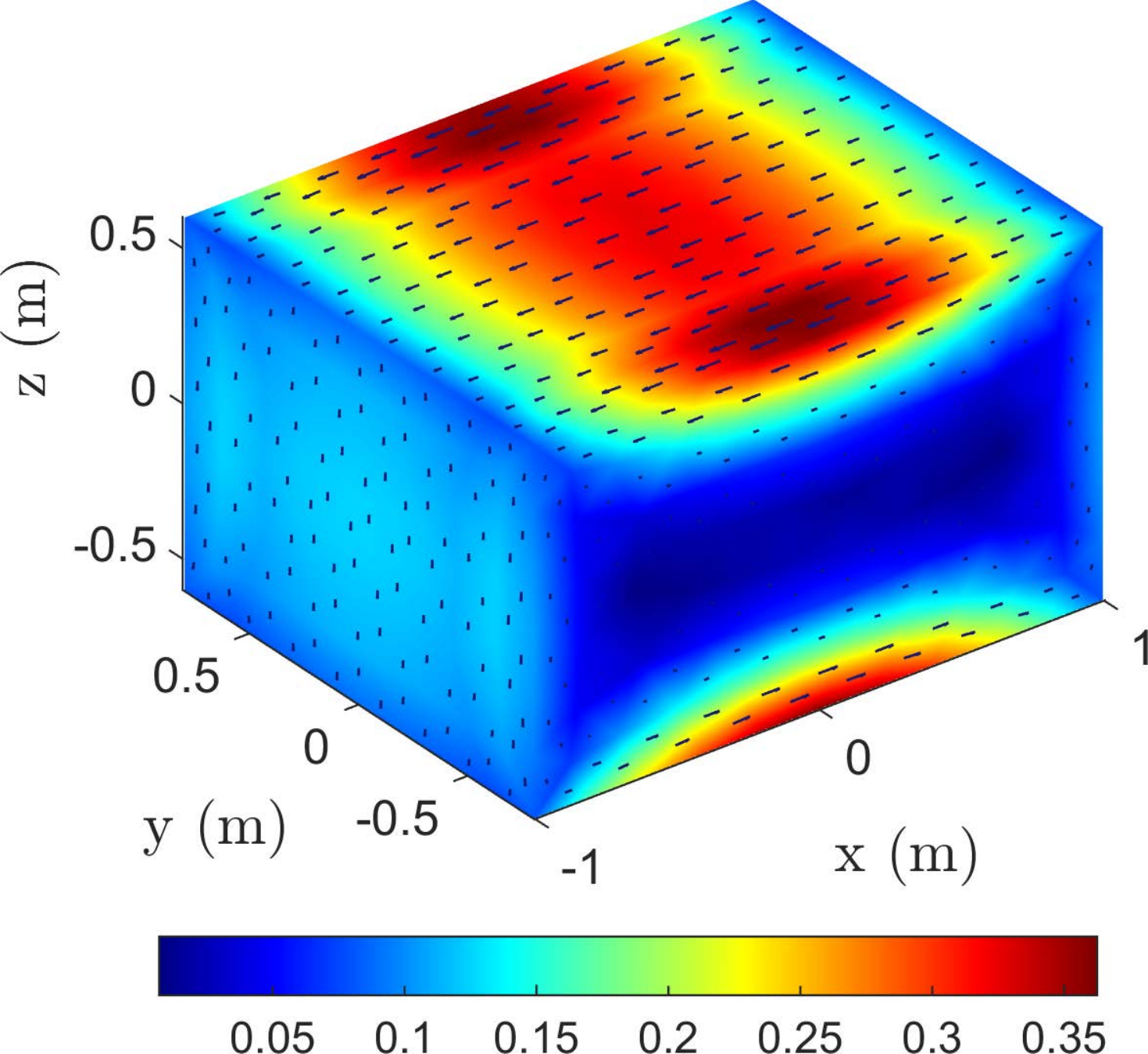} \label{cub-J3-119p88m-cfie}}
\subfigure[]{\includegraphics[scale=0.27]{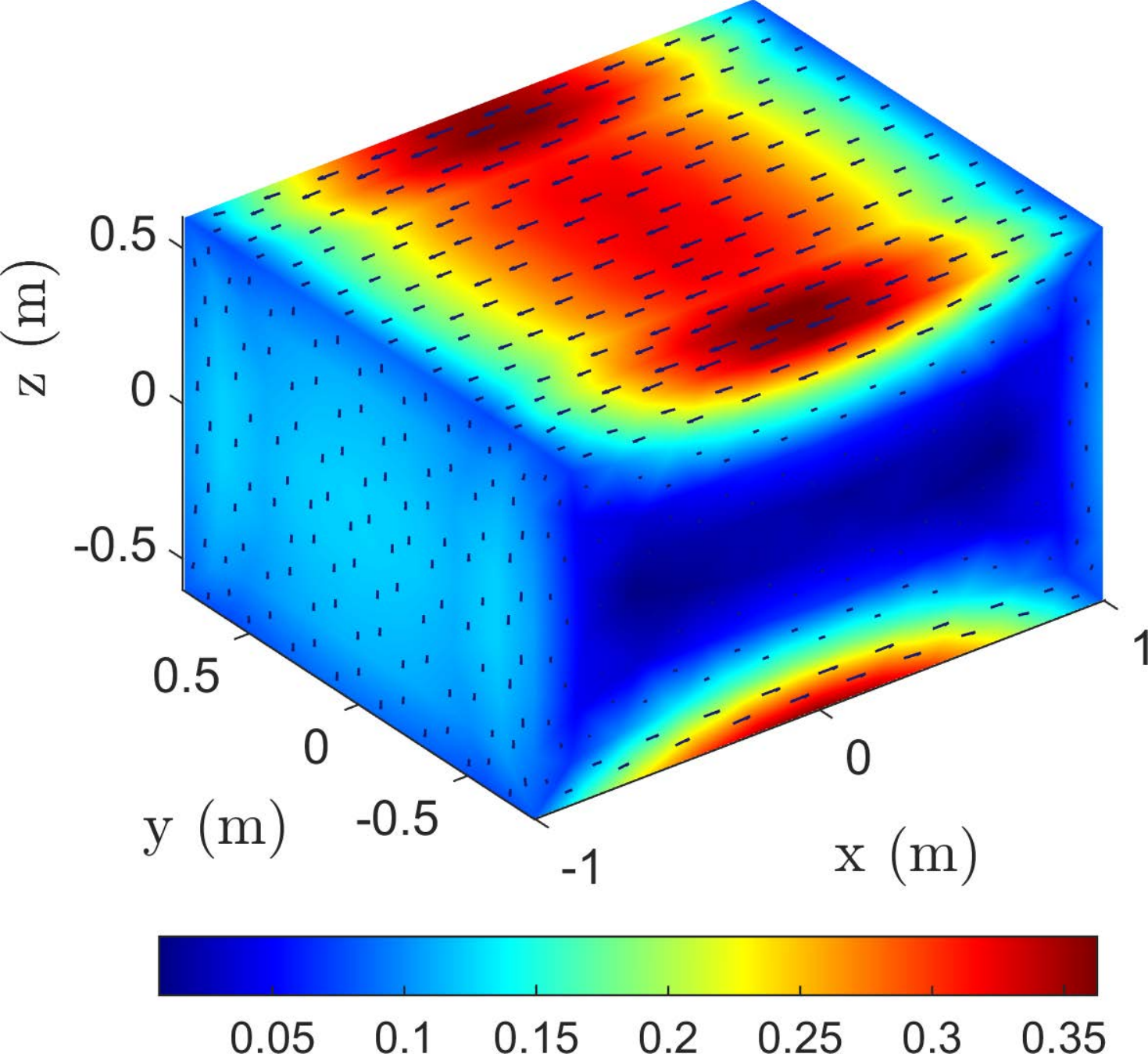} \label{cub-J3-119p88m-cmpcfie}}
\subfigure[]{\includegraphics[scale=0.27]{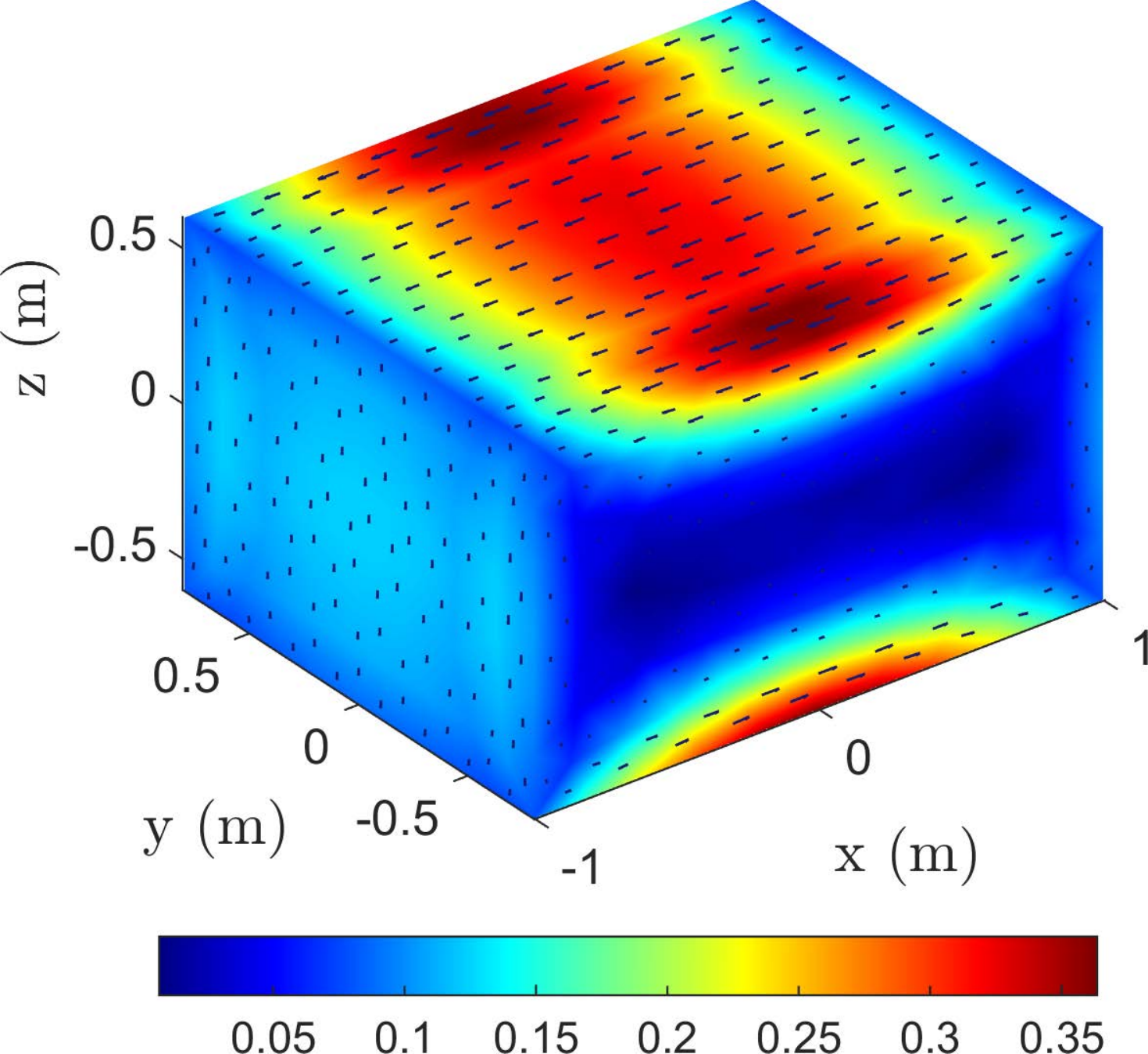} \label{cub-J3-121m-efie}}
\caption{Cuboid modal current $\mathbf{J}_3$: (a) Pattern inconsistent with others, computed by EFIE at $119.88$~MHz.  (b) Correct pattern computed by CFIE at $119.88$~MHz. (c) Correct pattern computed by CMP-CFIE at $119.88$~MHz. (d) Correct pattern computed by EFIE at $121$~MHz. The corresponding characteristic values of the $4$ modes are $0.4890$, $0.4892+0.0007i$, $0.4891+0.0041i$, and $0.4705$, respectively.} \label{cub-J3} 
\end{figure}

\begin{figure}[!t]
	\centering
	\subfigure[]{\includegraphics[scale=0.22]{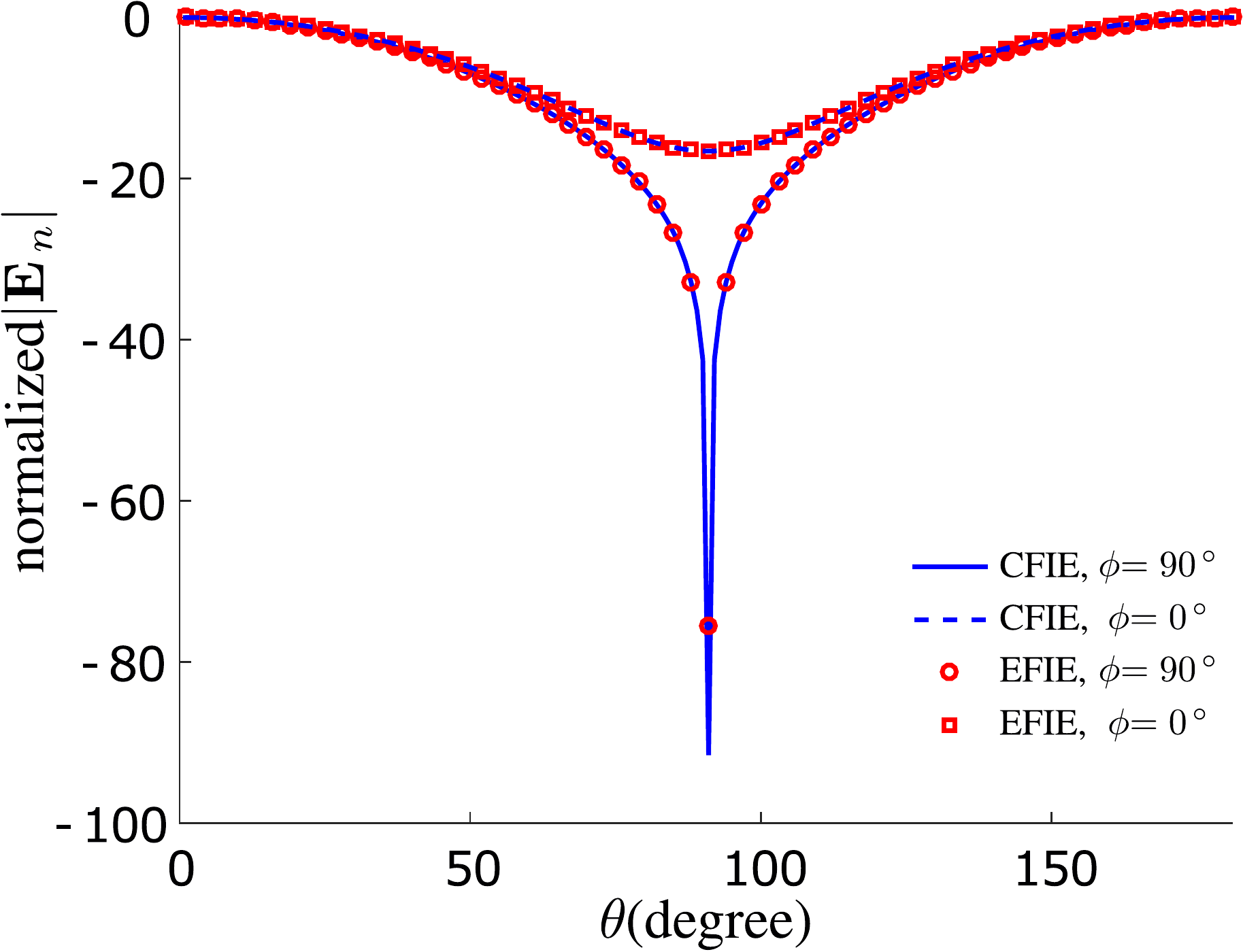} \label{J3-ff-ir}}
	\subfigure[]{\includegraphics[scale=0.22]{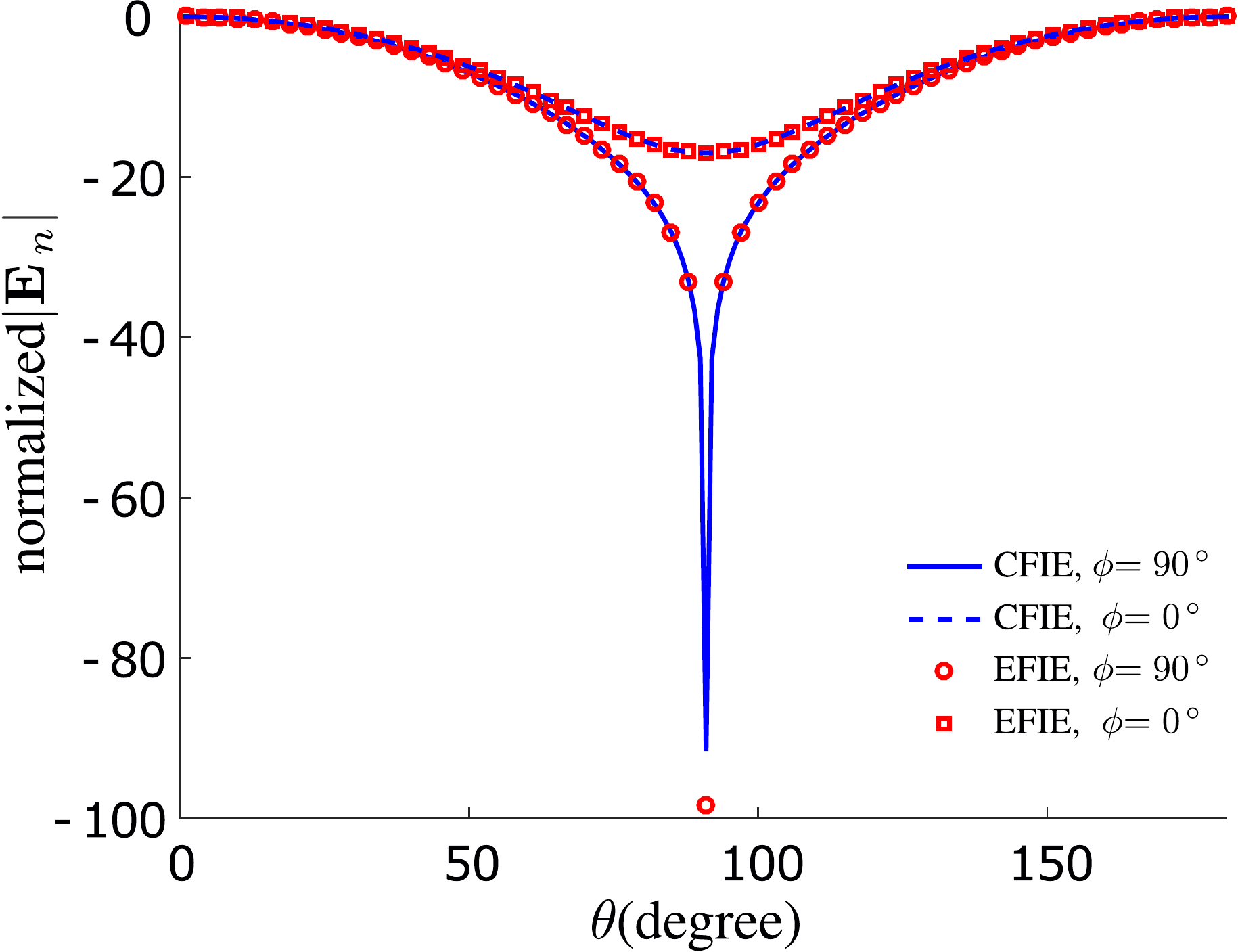} \label{J3-ff-nir}} 
	\caption{Far field patterns of cuboid mode $\mathbf{J}_3$: (a) at $119.88$~MHz. (b) at $121$~MHz. } \label{sph-J3_ff}
\end{figure}

At the internal resonance frequency $119.88$~MHz, we compute $50$ modes with the smallest characteristic values $\vert \lambda_n\vert$ by EFIE, CFIE, and CMP-CFIE, respectively, and list the first few characteristic values in Table~{\ref{cubCV}}. Good agreement is observed except for that $\lambda_n$ resulted from CFIE or CMP-CFIE are not purely real. This is due to the accuracy inconsistency between EFIE and MFIE solutions. 
Some characteristic currents cannot be correctly computed by the EFIE based TCM at spurious internal resonances. We demonstrate the current pattern of $\mathbf{J}_3$ as an example, which is shown in Fig.~\ref{cub-J3-119p88m-efie}. The correct current patterns can be obtained by CFIE and CMP-CFIE based TCMs, which are shown in Figs.~\ref{cub-J3-119p88m-cfie} and (c), respectively. When the operating frequency is shifted off the internal resonance one, the correct current mode can be found among the EFIE results after mode tracking, as shown in Fig.~\ref{cub-J3-121m-efie}. A comparison of the normalized far field patterns of this current at $119.88$~MHz and $121$~MHz are plotted in Fig.~\ref{J3-ff-ir} and (b), respectively, where good agreement between the results of EFIE and CFIE  is observed. Table~\ref{cubIterNum} shows $N_{out}$, $N_{in}$ and CPU times of different TCMs. It is obvious that both CFIE and CMP-CFIE based TCMs are more efficient than the EFIE based one. The speed-accuracy trade-off needs to be considered as setting $\alpha$ large suppresses the imaginary parts of $\lambda_n$ but increases the CPU time. Based on the results presented in \cite{Bagci09, Peeters}, the CMP-CFIE scheme is promising for large-scale applications where the conventional EFIE  fails to function.

\begin{table}[]
	\renewcommand{\arraystretch}{1.3}
	\caption{Characteristic Values (CVs) Computed by EFIE, CFIE, and CMP-CFIE at Internal Resonance $119.88$~MHz (Cuboid)}
	\vskip0.in
	\begin{center}
		\setlength{\tabcolsep}{1.em}
		\small \begin{tabular}{c c c c}\hline
			CV & EFIE & CFIE & CMP-CFIE \\ \hline
			$\lambda_1$ & $0.2389$ &$0.2391+0.0011i$  & $0.2390+0.0013i$  \\ 
			$\lambda_2$ & $-0.4643$ & $-0.4639+0.0022i$ & $-0.4637+0.0041i$  \\
			$\lambda_3$ & $0.4890$ & $0.4892+0.0007i$ & $0.4891+0.0010i$  \\
			$\lambda_4$ & $0.4979$ & $0.4980+0.0002i$ & $0.4979+0.0004i$  \\ 
			$\lambda_5$ & $-0.6834$ & $-0.6826+0.0038i$ & $-0.6823+0.0087i$  \\ \hline
		\end{tabular}
	\end{center}
	\label{cubCV}
\end{table}

\begin{table}[]
\renewcommand{\arraystretch}{1.3}
\caption{Iteration Numbers and CPU Times of EFIE, CFIE, and CMP-CFIE at Internal Resonance $119.88$~MHz (Cuboid)}
\vskip0.in
\begin{center}
\setlength{\tabcolsep}{1.2em}
\small \begin{tabular}{c c c c}\hline
TCM & EFIE & CFIE & CMP-CFIE \\ \hline
$N_{out}$ & $100$ & $100$ & $100$ \\ 
$N_{in}$ & $265.7$ & $192.3$ & $80.2$ \\ 
 CPU time (secs) & $240.0$ & $141.2$ & $44.2$ \\ \hline
\end{tabular}
\end{center}
\label{cubIterNum}
\end{table}

\begin{figure}[!t]
\centering
\subfigure[]{\includegraphics[scale=0.25]{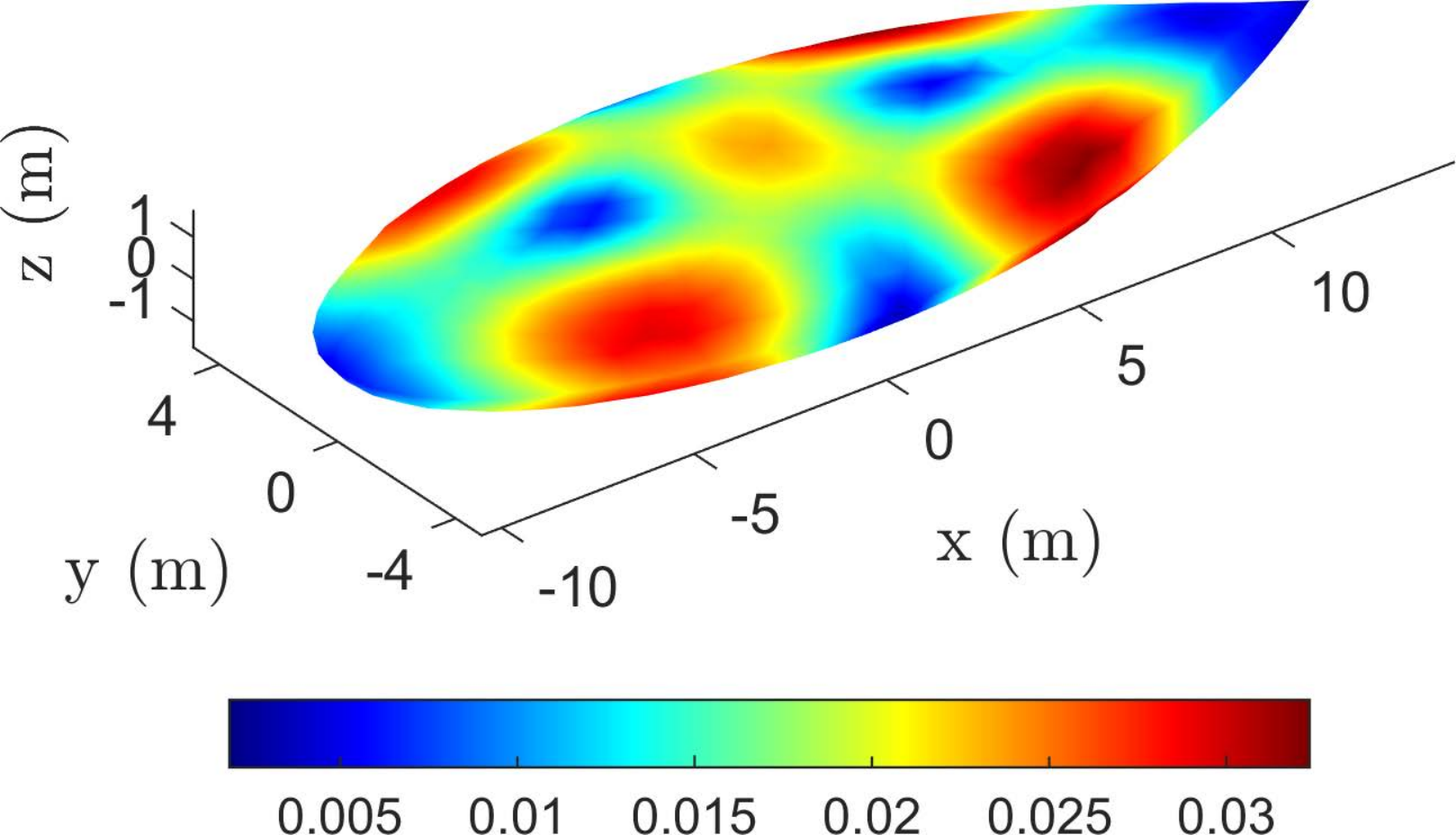} \label{almond-Jefie}}
\subfigure[]{\includegraphics[scale=0.25]{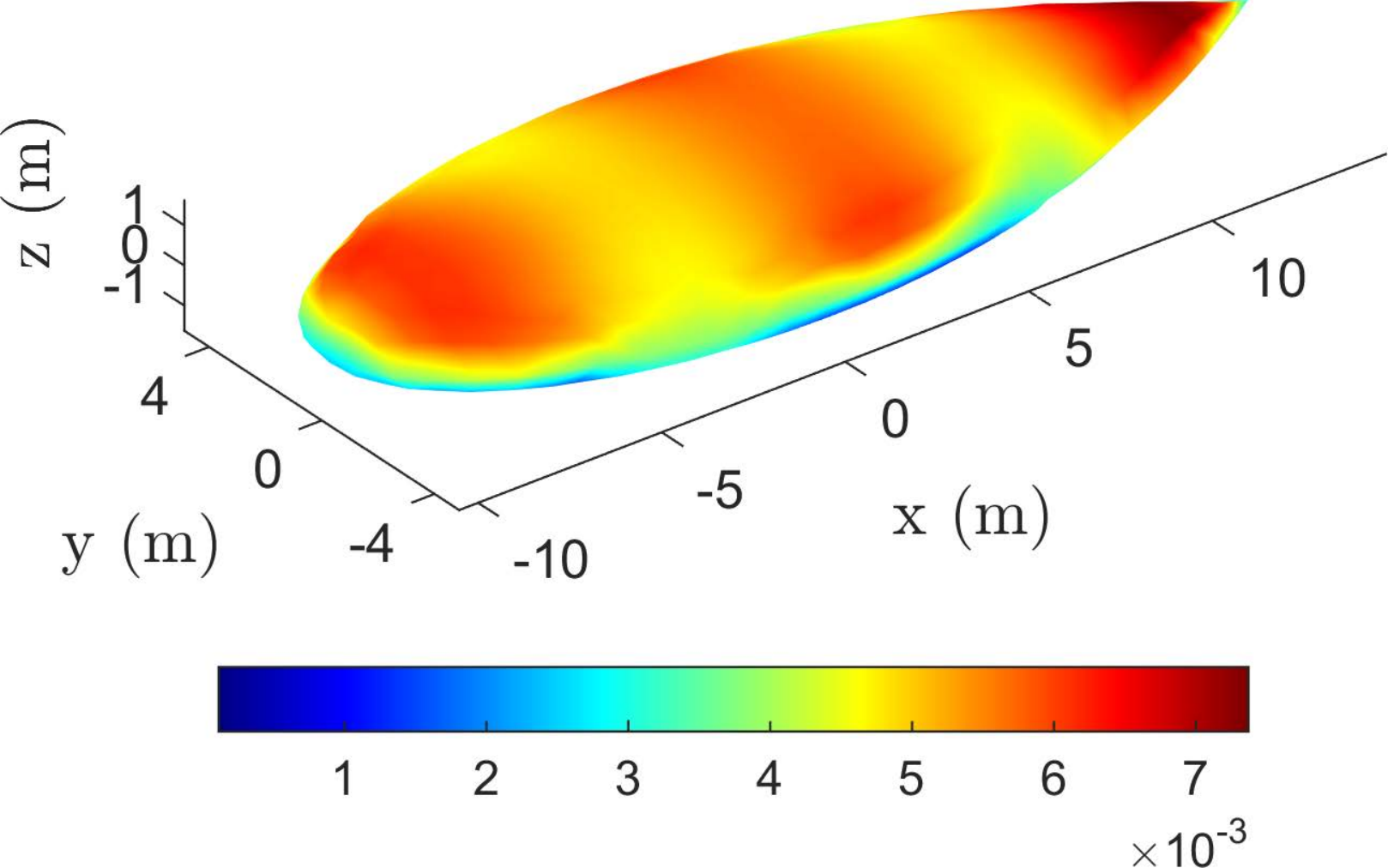} \label{almond-Jcfie}}
\subfigure[]{\includegraphics[scale=0.25]{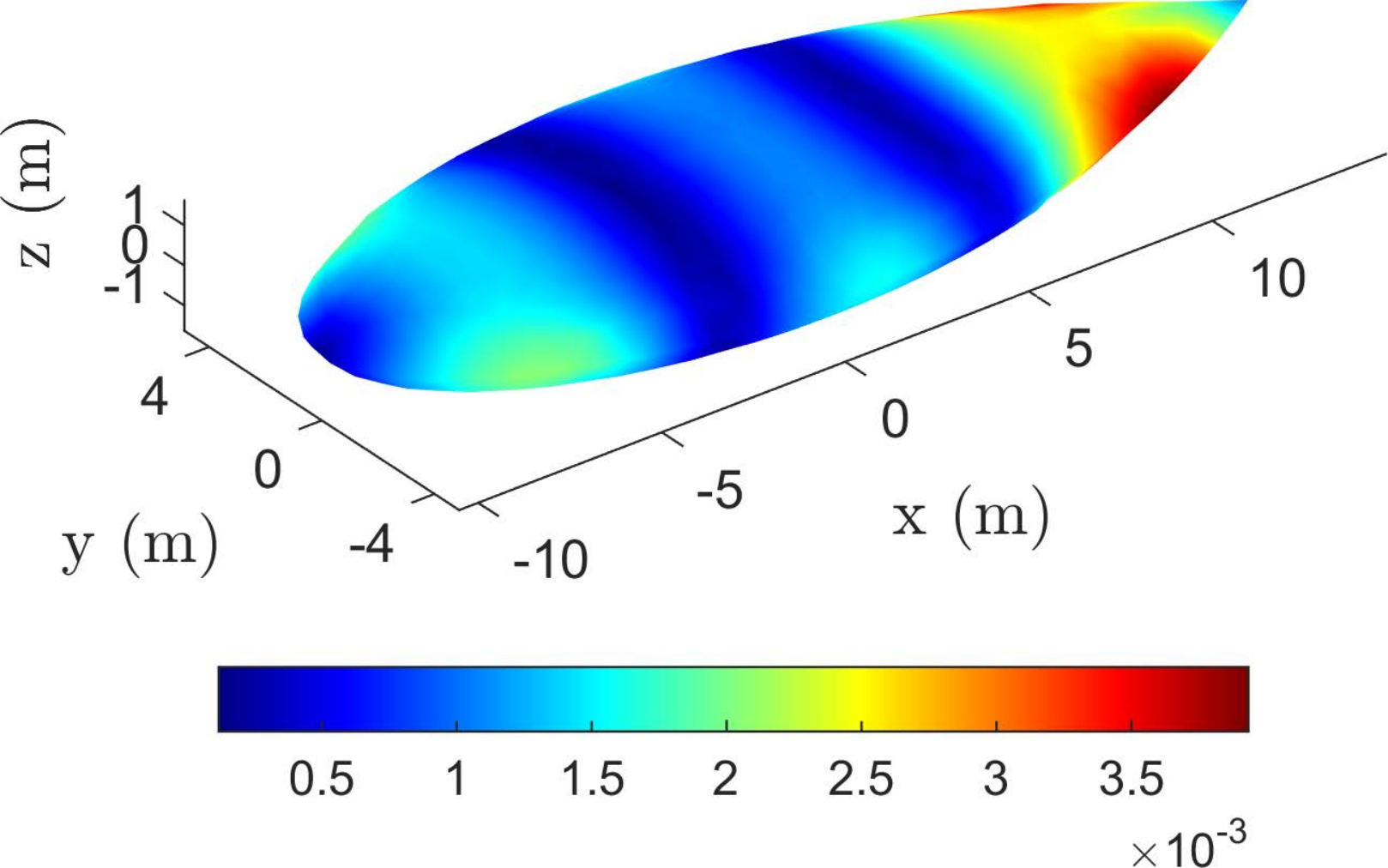} \label{almond-J5}}
\subfigure[]{\includegraphics[scale=0.25]{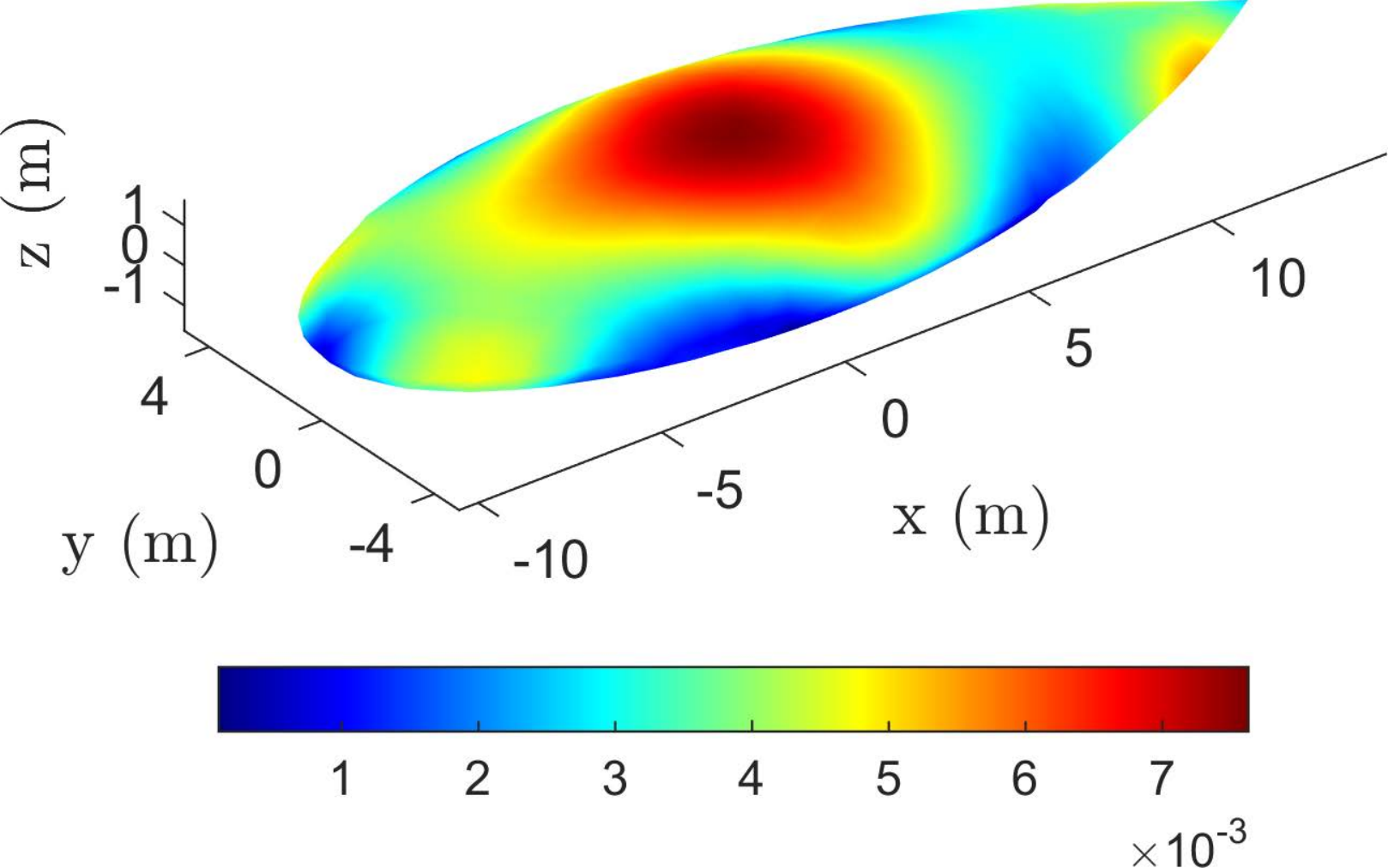} \label{almond-J15}}
\subfigure[]{\includegraphics[scale=0.25]{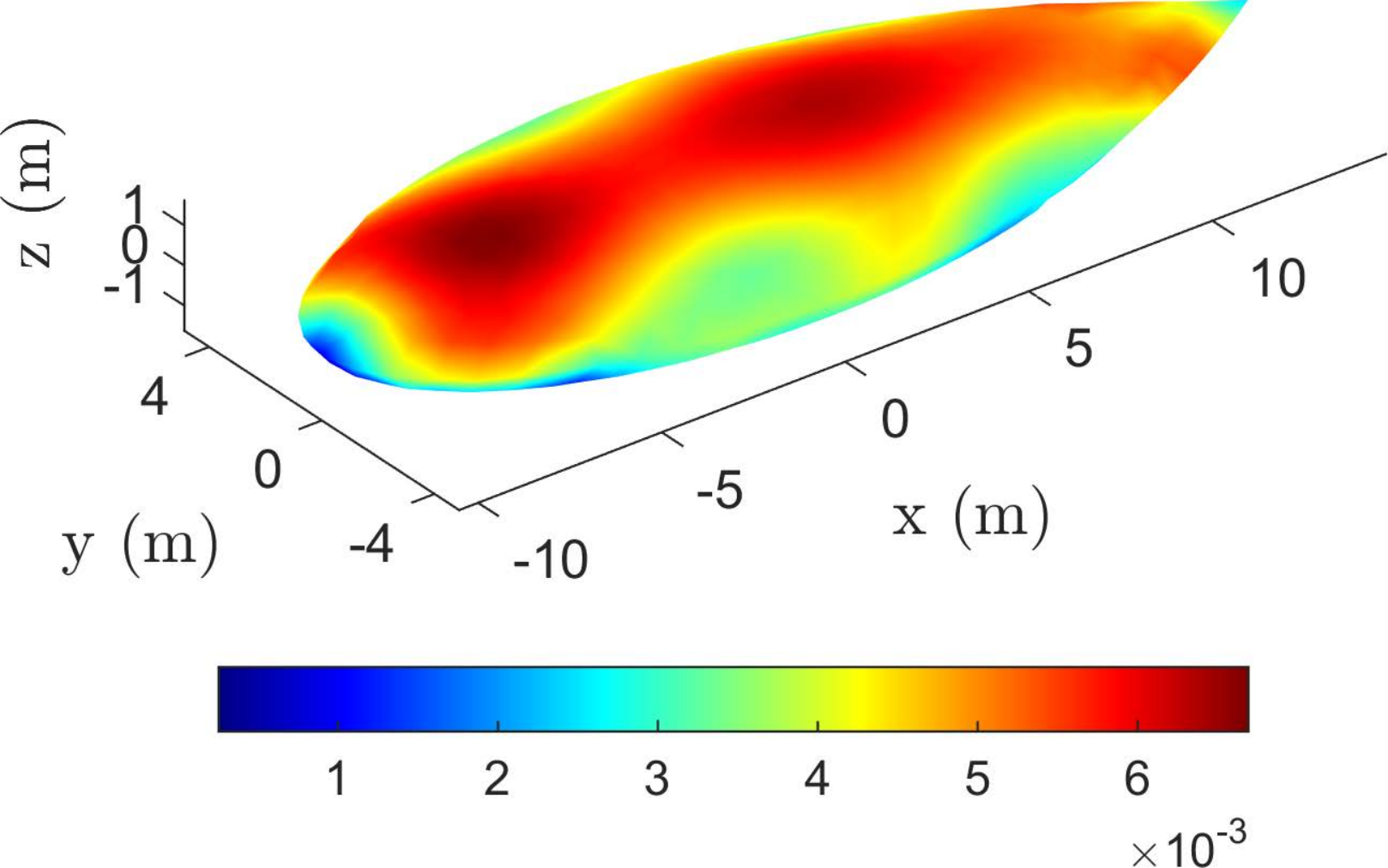} \label{almond-J25}}
\subfigure[]{\includegraphics[scale=0.25]{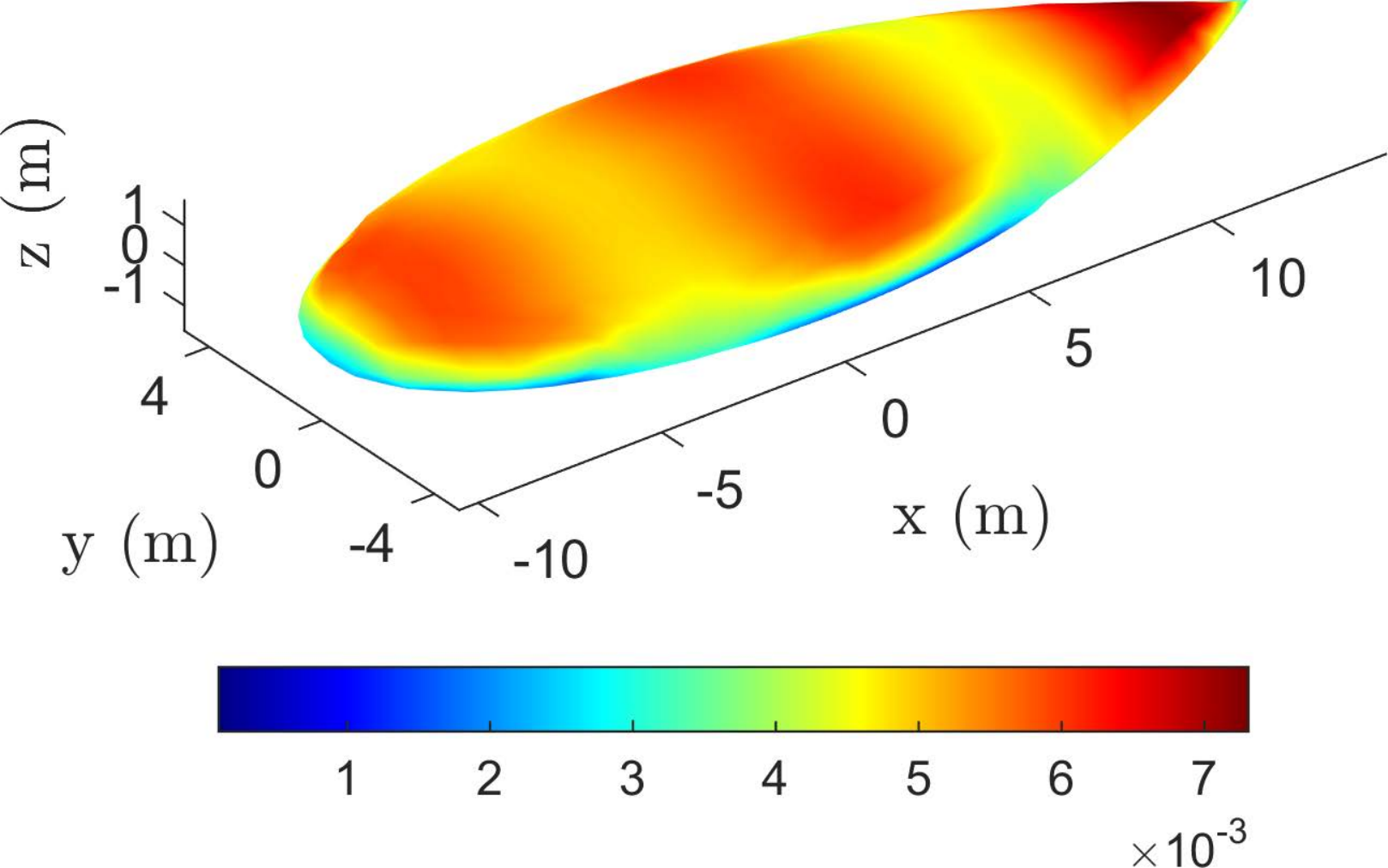} \label{almond-J75}}
\caption{Directly computed and reconstructed currents on a NASA almond at internal resonance $26.81$~MHz: (a) EFIE solution.  (b) CFIE solution. (c) Current reconstructed using $5$ modes. (d) Current reconstructed using $15$ modes. (e) Current reconstructed using $25$ modes. (f) Current reconstructed using $75$ modes. } \label{almond-J}
\end{figure}

In the last example, we demonstrate the characteristic mode expansion with a NASA almond where $1\,884$ edges are used. We first solve a scattering problem where the almond is illuminated by a $\hat x$-polarized plane wave propagating along $-\hat z$-direction at $26.81$~MHz. The induced current computed by EFIE is corrupted by an internal resonance mode, as shown in Fig.~\ref{almond-Jefie}. However, by  setting $\alpha$ to $0.5$, one can use CFIE to find the correct induced current as illustrated in Fig.~\ref{almond-Jcfie}. 

\begin{figure}[!t]
\centering
\includegraphics[scale=0.4]{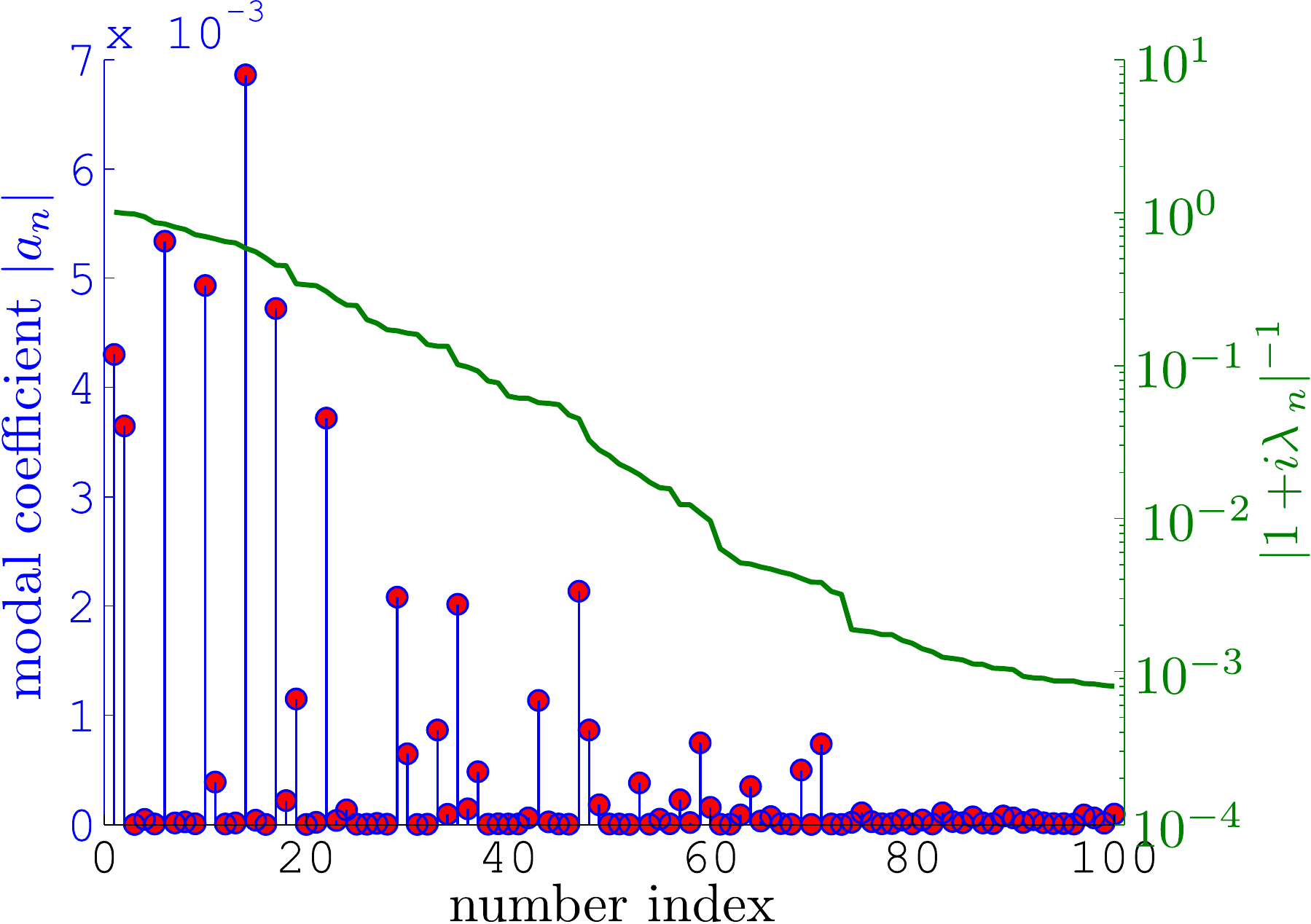}
\caption{Modal coefficients $\vert a_n\vert$ of the first $100$ characteristic modes.} \label{alm-an}
\end{figure}

\begin{figure}[!t]
\centering
\includegraphics[scale=0.4]{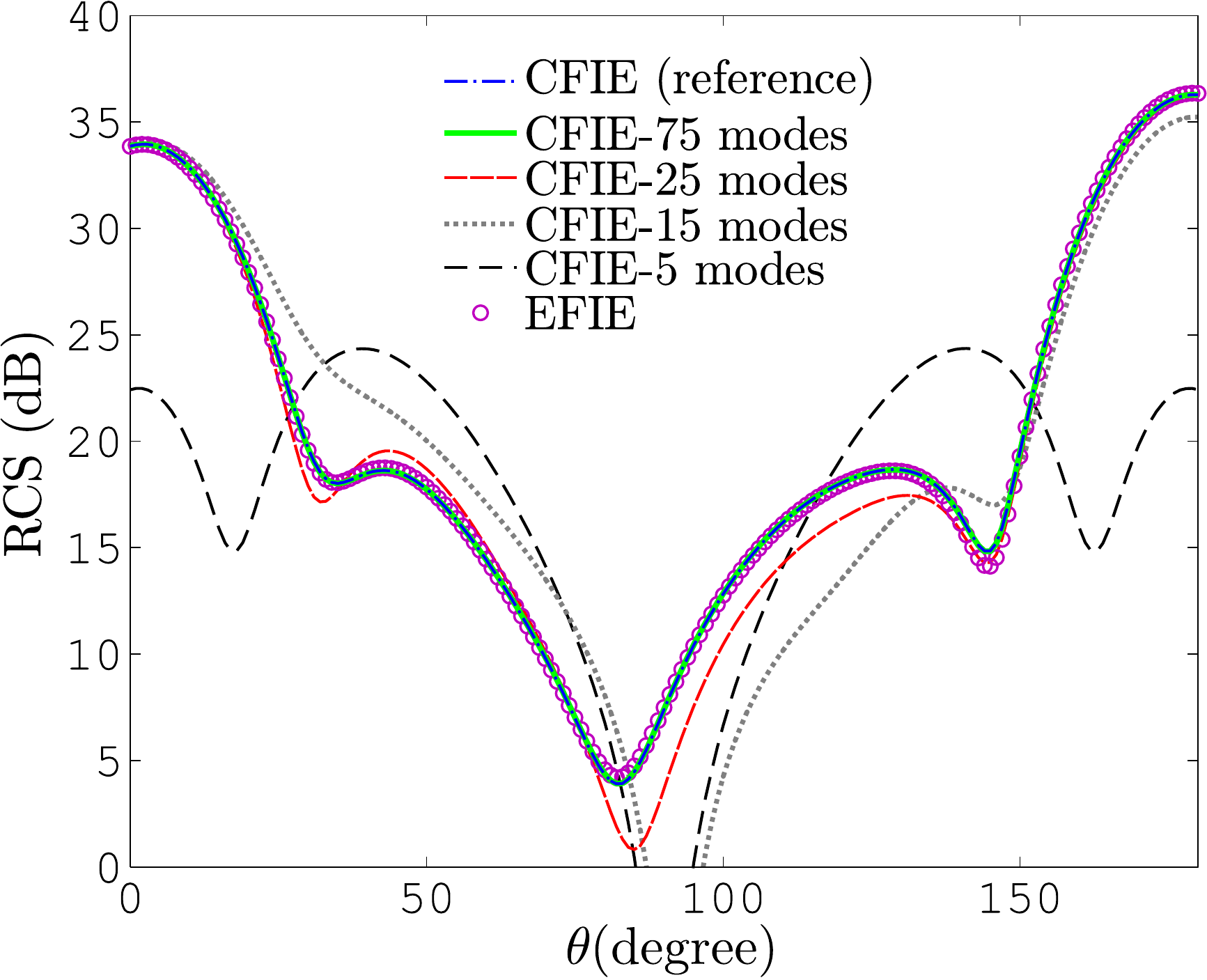}
\caption{RCSs of directly computed and reconstructed currents.} \label{alm-RCS}
\end{figure}

We find a few important characteristic modes with the CFIE based TCM, and use them to reconstruct the induced current. Thus, only a few columns of $\bar{\mathbf{J}}$ and $\bar{\mathbf{J}}^a$ in (\ref{cm-exp-coef}) which correspond to small $|1+i\lambda_n|$  are kept to compute the modal coefficients. The first $100$ modal coefficients $a_n$ are plotted in Fig.~\ref{alm-an}. It shows that not all modes with large $|1+i\lambda_n|^{-1}$ can be efficiently excited as their modal-source interaction $\left(\bar{\mathbf{J}}^a\right)^T\mathbf{F}_{inc}$ is negligible.  Figs.~\ref{almond-J5} to (f) illustrate the currents reconstructed using $5$, $15$, $25$ and $75$ characteristic modes, respectively. Figure~\ref{alm-RCS} plots RCSs ($\phi=0^\circ$) of the currents computed with EFIE, CFIE, as well as those reconstructed using different numbers of characteristic modes. It is obvious that the reconstructed current and the associated RCS converge to correct results as more modes are included. In this example, good agreement is observed when $75$ modes are considered. Hence, the system's order can be greatly reduced from $1\,884$ to $75$, which offers a model order reduction (MOR) based on characteristic mode expansion.

\section{Conclusion}
A CFIE based theory of characteristic mode (TCM) is presented in this paper to overcome the difficulty of slow convergence of EFIE solutions around frequencies of spurious internal resonances. Since the MFIE based formulation shares a common set of non-trivial characteristic pairs with the EFIE based one, they are combined to form a generalized eigenvalue problem which can be easily casted into a standard one where the spurious internal resonance corruption is removed. A CMP-CFIE based TCM is further formulated to enhance the performance of the proposed scheme. MOR based on the characteristic mode expansion is also presented which may serve as a useful tool in system design and optimization. The validity and efficiency of the proposed scheme are demonstrated in a few numerical experiments. It may be incorporated with fast algorithms such as MLFMA to compute characteristic modes of large-scale geometries.

\ifCLASSOPTIONcaptionsoff
  \newpage
\fi

\bibliographystyle{IEEEtran}
\bibliography{thesisrefs}







\end{document}